\documentclass[a4paper, 11pt]{article}   
\usepackage[dvipsnames]{xcolor}
\usepackage{authblk}
\usepackage{graphicx}		
\usepackage{amssymb}
\usepackage{amsmath}
\usepackage{algorithmic}
\usepackage{algorithm}
\usepackage{color}
\usepackage{tikz}
\usetikzlibrary{positioning, calc}
\usepackage[toc,title]{appendix}
\usepackage{array, multirow, booktabs}
\usepackage[nameinlink]{cleveref}

\newcommand{\Real}{\mathop{\mathrm{Re}}}

\newcommand{\bx}{\mathbf{x}}

\newcommand{\Cov}{\mathbb{C}\text{ov}}
\newcommand{\Var}{\mathbb{V}\text{ar}}
\newcommand{\Exp}{\mathbb{E}}
\title{Emulating complex dynamical simulators with random Fourier features}
\author[1]{Hossein Mohammadi\thanks{Corresponding author: h.mohammadi@exeter.ac.uk}}
\author[1]{Peter Challenor}
\author[1, 2]{Marc Goodfellow}
\affil[1]{\footnotesize Faculty of Environment, Science and Economy, University of Exeter, Exeter, UK}
\affil[2]{\footnotesize Living Systems Institute, University of Exeter, Exeter, UK}
\date{}	
\begin{document}
\maketitle
%===================================================================================================
\begin{abstract}
A Gaussian process (GP)-based methodology is proposed to emulate complex dynamical computer models (or simulators). The method relies on emulating the numerical flow map of the system over an initial (short) time step, where the flow map is a function that describes the evolution of the system from an initial condition to a subsequent value at the next time step. This yields a probabilistic distribution over the entire flow map function, with each draw offering an approximation to the flow map. The model output times series is then predicted (under the Markov assumption) by drawing a sample from the emulated flow map (i.e., its posterior distribution) and using it to iterate from the initial condition ahead in time. Repeating this procedure with multiple such draws creates a distribution over the time series. The mean and variance of this distribution at a specific time point serve as the model output prediction and the associated uncertainty, respectively. However, drawing a GP posterior sample that represents the underlying function across its entire domain is computationally infeasible, given the infinite-dimensional nature of this object. To overcome this limitation, one can generate such a sample in an approximate manner using random Fourier features (RFF). RFF is an efficient technique for approximating the kernel and generating GP samples, offering both computational efficiency and theoretical guarantees. The proposed method is applied to emulate several dynamic nonlinear simulators including the well-known Lorenz and van der Pol models. The results suggest that our approach has a promising predictive performance and the associated uncertainty can capture the dynamics of the system appropriately. 
\end{abstract}
\section{Introduction}
\label{sec:introduction}
%===================================================================================================
The problem of predicting the output of complex computer codes (or simulators) occurs frequently in many real-world applications \cite{jia2013, wang2019}. Such simulators are based on complicated mathematical equations and can be computationally intensive. Hence the number of simulation runs is limited by our budget for computation. One way to overcome this problem is to create a surrogate of the computer code which is cheap-to-evaluate. Surrogates are statistical representation of the true model and are constructed based on a limited number of simulation runs. A survey of the most widely used surrogate models is presented in \cite{alizadeh2020, jin2001}. Among broad types of surrogate models, Gaussian process (GP) emulators \cite{GPML} have become the gold standard in the field of the design and analysis of computer experiments \cite{chen2006, sacks1989, santner2003}. This is due to their statistical properties such as the flexibility and computational tractability, see \Cref{sec:GPR} for further detail, though other methods from machine learning such as neural networks are also used \cite{wang2019}.

In this work, we focus on the emulation of deterministic \emph{dynamical simulators} which are based on a set of ordinary differential equations (ODE). Dynamical simulators are an important class of computer codes whose output is a time series. Here, the whole time series is generated in a single run of the simulator, given initial conditions. This can be regarded as a simple, one-step simulation being run iteratively for many time steps \cite{conti2009}. A dynamic computer code simulates the evolution of a physical process over time and is widely used in various fields including biology \cite{raue2013} and climate science \cite{roberts2018}. The Hindmarsh-Rose (HR) model \cite{hindmarsh1984}, which simulates the dynamics of a single neuron, is a specific example that is discussed in more detail in \Cref{sec:hindmarsh}. 

Emulating dynamical computer codes is an active field of research and has been tackled via different statistical and machine learning approaches, see e.g. \cite{brunton2019, castelletti2012, chen2018, mohammadi2019, raissi2019}. One may consider this problem as a special case of multi-output GPs \cite{alvarez2011, conti2010, fricker2013} with a temporal dependency between the observations. However, the size of the output dimension in dynamical models is usually too high to be treated through multi-output GPs. To address this issue, one can first apply techniques such as principal component analysis \cite{higdon2008, jia2013} or wavelet decomposition \cite{bayarri2007, dernoncourt2015} to reduce the output dimensionality. The pitfall is that we lose some information by not keeping all the components. Another approach is proposed in \cite{kennedy2001} which accounts for time using an extra input parameter. This method increases the computational complexity and is reported to be inefficient \cite{conti2009, machac2016}. The idea of forecasting the time series through iterative one-step ahead predictions is developed in \cite{bhattacharya2007, conti2009}. This method relies on emulating the transition function from one time point to the next, under the assumption that the model output at time $t$ depends only on the observation at time $t - 1$, i.e. the Markov property. The work is continued in \cite{mohammadi2019} considering the input uncertainty at each step of the iterative prediction process. 

This paper presents a novel data-driven approach for emulating complex dynamical simulators relying on emulating the numerical flow map over a short period of time. The flow map is a function that maps an initial condition to the solution of the system at a future time $t$. We emulate the numerical flow map of the system over the initial (short) time step via GPs. The idea is that GP emulators model the underlying function (in this case, the flow map) as a probabilistic distribution, and their sample paths provide a characterisation of the function throughout its entire domain. These sample paths extend the notion of merely being a distribution over individual function values at specific points, such as those generated from a multivariate normal distribution. The model output time series is then predicted relying on the Markov assumption; a sample path from the emulated flow map is drawn and employed in an iterative manner to perform one-step ahead predictions. By repeating this procedure with multiple draws, we acquire a distribution over the time series whose mean and variance at a specific time point serve as the model output prediction and the associated uncertainty, respectively. However, obtaining a GP sample path, evaluable at any location in the domain for use in one-step ahead predictions, is infeasible. To address this challenge, we employ RFF \cite{rahimi2008}, as described in \Cref{sec:kernel_approx}. RFF is a technique for approximating the GP kernel using a finite number of its Fourier features. The resulting approximate GP samples, generated with RFF, are analytically tractable, providing both theoretical guarantees and computational efficiency.

The rest of the paper is organised as follows. First, we give a brief overview of dynamical systems. Then, in \Cref{sec:GPR} we introduce GP emulators. \Cref{sec:kernel_approx} reviews RFF and its application to kernel approximation and GPs. Section \Cref{sec:Emul_Dynam_Sys} describes our proposed method for emulating dynamical models. Numerical results are provided in \Cref{sec:Num_Res} where we apply our method to emulate several dynamical systems. Finally, \Cref{sec:conclusion} presents our conclusions.
%===========================================================================
\paragraph*{Dynamical system}
%===========================================================================
A dynamical system represents the evolution of a phenomenon over time according to a fixed mathematical rule. Here we focus on continuous time systems represented by a set of ordinary differential equations, hence solutions of these equations are given by the vector of \emph{state variables}, $\bx(t) = (x_1(t), \ldots, x_d(t))^\top$ which determines the state of the system at time $t \in \mathbb{R}$. The space that consists of all possible values of the state variables is called the \emph{state (phase) space} denoted by $\mathcal{X} \subset \mathbb{R}^d$. The ordinary differential equations define a \emph{vector field}, $\mathbf v$, in $\mathbb{R}^d$, that is tangent to the solution \cite{strogatz2007}, i.e.
\begin{equation}
	\mathbf  v : \mathcal{X} \mapsto \mathcal{X} ~ , \quad \frac{d}{dt} \bx(t) = \mathbf  v \left(\bx(t) \right) .
	\label{vector_field}
\end{equation}
We assume that the system is autonomous, meaning that the associated vector field does not depend on time explicitly. 

Let $\bx_0$ be an initial condition, which represents the state of the system at an initial time $t_0$, and $\Delta t$ a time step. The \emph{flow map} ($\mathbf F$) is a function that maps $\bx_0$ to the state at the next time step $t_1 = t_0 + \Delta t$ \cite{cvitanovic2016}:
\begin{equation}
	\mathbf F : \mathcal{X} \times \mathbb{R}  \mapsto \mathcal{X}, \quad \mathbf F\left(\bx_0, \Delta t\right) = \bx\left(t_1\right) .
	\label{flow_map}
\end{equation}
The flow map can be interpreted as mapping the initial condition to the end point of a short trajectory traced out in the state space from $\bx_0$ to $\bx(t_1)$. In this work, we assume that the time step $\Delta t$ is fixed, and hence the flow map is a function of $\bx_0$ only. It is important to note that in the proposed method, the GP emulator is built over the initial time step $\Delta t = t_1 - t_0$. Then, the GP samples are employed in an iterative manner to perform one-step ahead predictions, and predict the whole time series, see \Cref{sec:Emul_Dynam_Sys} for further details.
%===================================================================================================
\section{Gaussian process emulators}
\label{sec:GPR}
%===================================================================================================
Our aim is to build statistical representations of the flow map based on GPs to enable efficient construction of trajectories with quantified uncertainty. The potential benefits of this approach are shown in \cite{mohammadi2019}. Before explaining the advances in the current manuscript, we introduce GP emulators. In this section, the notation $\bx$ is used for general input parameters. However, the vector of initial conditions $\bx_0$ are the inputs in the problem of interest that is the numerical flow map emulation.

Let $f(\bx), \bx \in \mathcal{X}$, be the function we wish to emulate. We do this using the stochastic Gaussian process $Y(\bx)$ given by
\begin{equation}
	Y(\bx)= \mu(\bx) + Z(\bx) ,
	\label{stochastic_process}
\end{equation}
in which $\mu: \mathcal{X} \mapsto \mathbb{R}$ is the \emph{trend function} that can take any functional form. In this work, the trend function is linear
$\mu(\bx) = \mathbf{q}(\bx)^\top \boldsymbol{\beta}$ with $\mathbf{q}(\bx) = \lbrack q_1(\bx), \ldots, q_r(\bx) \rbrack^\top$ and $\boldsymbol{\beta} =  \lbrack \beta_1, \ldots, \beta_r \rbrack^\top$ being the vector of basis (regression) functions and the corresponding coefficients, respectively. The second component in \Cref{stochastic_process}, $Z(\bx)$, is a centred (or zero mean) GP with the covariance function $\Cov\left(Z(\bx), Z(\bx^\prime)\right) = \sigma^2k \left(\bx, \bx^\prime\right) ,~\forall \bx, \bx^\prime \in\mathcal{X}$. The scalar $\sigma^2$ is the \emph{process (signal) variance} and controls the scale of the amplitude of $Z(\bx)$. The function $k : \mathcal{X} \times \mathcal{X} \mapsto \mathbb{R}$ is the kernel or correlation function and regulates the level of smoothness of $Z(\bx)$. A kernel is called \emph{stationary (shift invariant)} if it depends only on the difference between its inputs: $k(\bx, \bx^\prime) = k(\bx - \bx^\prime).$ One of the most common stationary correlation functions is the squared exponential (SE) kernel defined as \cite{GPML}
\begin{equation}
	k_{SE}(\bx, \bx^\prime) = \exp \left( -0.5 \left( \bx -\bx^\prime \right)^\top \boldsymbol{\Delta}^{-2}  \left(  \bx -\bx^\prime \right) \right) ,
	\label{SE_kernel}
\end{equation}
where the diagonal matrix $\boldsymbol{\Delta} \in \mathbb{R}^{d\times d}$ consists of the \emph{correlation length-scales}. They are denoted by the vector $\boldsymbol{\delta} = \lbrack \delta_1, \ldots, \delta_d \rbrack^\top$. 

Let $\mathcal{D} = \lbrace \mathbf{X}, \mathbf{y} \rbrace$ denote the training data set in which $\mathbf{X} = \lbrack\bx^1, \ldots, \bx^n\rbrack^\top$ is called the \emph{design matrix} and includes $n$ points in the input space. The vector $\mathbf{y} = \lbrack f(\bx^1), \ldots, f(\bx^n) \rbrack^\top$ comprises the outputs at those locations. For our purpose, $\bx^1, \ldots, \bx^n $ represent samples taken from the state space of the system under study. Furthermore, in our method, the flow map function is emulated using data from the initial time step $\Delta t = t_1 - t_0$, as detailed in \Cref{sec:Emul_Dynam_Sys}. Typically, GPs are presented from a \emph{function space} perspective where the predictive distribution relies on the posterior $Y(\bx) \mid \mathcal{D}$. The GP predictive (posterior) mean and variance are
\begin{align}
	\label{GP_mean}
	&\Exp \lbrack Y(\bx) \mid \mathcal{D} \rbrack = \mu(\bx) +  \mathbf{k}(\bx)^\top \mathbf{K}^{-1} \left( \mathbf{y} - \boldsymbol{\mu} \right) ,	\\
	&\Var \left(Y(\bx) \mid \mathcal{D}  \right) = \sigma^2 \left( 1 - \mathbf{k}(\bx)^\top \mathbf{K}^{-1} \mathbf{k}(\bx) \right) ,
	\label{GP_var}
\end{align}
in which $\boldsymbol{\mu} = \mu(\mathbf{X})$, and $\mathbf{k}(\bx) = \left \lbrack k(\bx^1, \bx), \ldots, k(\bx^n, \bx) \right \rbrack^\top$ and $\mathbf{K}$ is an $n \times n$ correlation matrix whose $ij$-th element is $\mathbf{K}_{ij} = k(\bx^i, \bx^j) , \, \forall \bx^i, \bx^j \in \mathbf{X}$.  
%===========================================================================
\paragraph*{Parameter estimation}
%===========================================================================
The stochastic process $Y(\bx)$ depends on a set of parameters $\boldsymbol{\theta}  = \lbrace \sigma^2, \boldsymbol{\beta},   \boldsymbol{\delta} \rbrace$ that are generally unknown and need to be estimated from the data. In this work, we use the maximum likelihood method to estimate them. The logarithm of the likelihood function is
\begin{equation}
	\mathcal{L}(\boldsymbol{\theta} \mid \mathbf{y}) = - \frac{n}{2}\ln(2\pi\sigma^2) - \frac{1}{2}\ln(\mid\mathbf{K}\mid) - \frac{1}{2\sigma^2}  \left(\mathbf{y} - \boldsymbol{\mu} \right)^\top \mathbf{K}^{-1}  \left(\mathbf{y} - \boldsymbol{\mu} \right) ,
	\label{log_lik}
\end{equation}
where the correlation matrix $\mathbf{K}$ depends on $\boldsymbol{\delta}$ and $\boldsymbol{\mu} = \mathbf{Q}\boldsymbol{\beta} $ in which $ \mathbf{Q} = \lbrack\mathbf{q}(\bx^1), \ldots, \mathbf{q}(\bx^n) \rbrack^\top$. It is an $n\times r$ matrix called the \emph{experimental matrix} and comprises the evaluation of the regression functions at the training data. An estimate of $\boldsymbol{\beta}$ and $\sigma^2$ is obtained by taking the derivatives of $\mathcal{L}(\boldsymbol{\theta}\mid \mathbf{y})$ with respect to those parameters and setting the derivatives to zero. The estimated parameters have closed-form expressions given by
\begin{align}
	\label{beta_hat}
	\hat{\boldsymbol{\beta}} &= \left( \mathbf{Q}^\top \mathbf{K}^{-1} \mathbf{Q} \right)^{-1} \mathbf{Q}^\top \mathbf{K}^{-1} \mathbf{y} , \\
	\hat{\sigma}^2 &= \frac{1}{n} \left( \mathbf{y} - \mathbf{Q}\hat{\boldsymbol{\beta}} \right)^\top \mathbf{K}^{-1}  \left( \mathbf{y} - \mathbf{Q}\hat{\boldsymbol{\beta}} \right) .
	\label{sigma2_hat}
\end{align}
If the parameters $\boldsymbol{\beta}$ and $\sigma^2$ in (\Cref{log_lik}) are substituted with their estimates $\hat{\boldsymbol{\beta}}$ and $\hat{\sigma}^2$, the profile log-likelihood (after dropping the constants) is achieved as 
\begin{equation}
	\mathcal{L}_p(\boldsymbol{\delta} \mid \mathbf{y}) = - \frac{n}{2}\ln(\hat{\sigma}^2) - \frac{1}{2}\ln(\mid\mathbf{K}\mid) .
	\label{profile_log_lik}
\end{equation}
Finally, the length-scales can be estimated by solving the optimisation problem below
\begin{equation}
	\hat{\boldsymbol{\delta}} = \underset{\boldsymbol{\delta}}{\arg \max}  ~ \mathcal{L}_p(\boldsymbol{\delta}\mid \mathbf{y})  .
\end{equation}
%===================================================================================================
\section{Sampling from GP posteriors}
\label{sec:kernel_approx}
%===================================================================================================
This section provides the material necessary for sampling from the GP posterior distribution using RFF. To predict the model output time series, sample paths from the emulated flow map are drawn and employed in an iterative fashion for one-step ahead predictions. In this framework, the GP sample paths need to effectively represent the flow map function across its entire domain. However, obtaining a GP sample path that can be evaluated at any location $x\in \mathcal X$ in closed form is not possible \cite{kuss2006, bradford2018}. To overcome this issue, we employ RFF which is a popular technique for approximating the kernel and generating GP samples in an approximate manner, leveraging both theoretical guarantees and computational efficiency. The resulting approximate GP sample paths are analytically tractable. The other applications of kernel approximation with RFF can be found in Bayesian optimisation (which is referred to as Thompson sampling) \cite{shahriari2016}, deep learning \cite{mehrkanoon2018}, and big data modelling \cite{haitao2020}. We start the discussion by introducing RFF which offers an effective way to approximate stationary kernels.
%==============================================================
%==================================================================
\subsection{Kernel approximation with RFF} 
\label{sec:RFF}
%==================================================================
The covariance function $k$ can be expressed as an inner product $k\left(\bx, \bx^\prime\right) = \left\langle \phi(\bx), \phi(\bx^\prime) \right\rangle_{\mathcal{H}}$ where $\phi(\bx) : \mathcal{X} \mapsto \mathcal{H}$ is called the \emph{feature map}. It transforms the original space $\mathcal{X}$ into a higher (or infinite) dimensional reproducing kernel Hilbert space (RKHS) $\mathcal{H}$ \cite{scholkopf2001, hofmann2008}. A brief overview of RKHSs is provided in Appendix \ref{appendix:rkhs}. According to Bochner's theorem \cite{bochner1959}, the Fourier transform of the stationary kernel $k$ is
\begin{equation}
	k(\bx , \bx^\prime) = \int e^{-i \boldsymbol{\omega}^\top (\bx - \bx^\prime)} d \mathbb{P}(\boldsymbol{\omega}) ,
	\label{fourier_trans}
\end{equation}
where $\mathbb{P}(\boldsymbol{\omega})$ (the Fourier dual of $k$) is referred to as the \emph{spectral distribution} of the kernel. $\mathbb{P}(\boldsymbol{\omega})$ has all the properties of a cumulative distribution function except that $\mathbb{P}(\infty) - \mathbb{P}(-\infty)= k(\mathbf{0})$ needs not to be equal to one \cite{lindgren2006}. However, in classic correlation functions such as the SE kernel, $\mathbb{P}(\boldsymbol{\omega})$ is a proper cumulative distribution function because $k(\mathbf{0}) = 1$. In this situation, $p(\boldsymbol{\omega}) = \frac{d \mathbb{P}(\boldsymbol{\omega})}{d \boldsymbol{\omega}}$ is the density function of $\boldsymbol{\omega}$ and \Cref{fourier_trans} can be rewritten as \cite{henrandez-lobato2014}
\begin{align}
	\nonumber k(\bx , \bx^\prime) = \int e^{-i \boldsymbol{\omega}^\top (\bx - \bx^\prime)} p(\boldsymbol{\omega}) d \boldsymbol{\omega} & =  \Exp_{p(\boldsymbol{\omega})} \left\lbrack e^{-i \boldsymbol{\omega}^\top (\bx - \bx^\prime)} \right\rbrack \\
	\nonumber & = \Exp_{p(\boldsymbol{\omega})} \left\lbrack \Real \left( e^{-i \boldsymbol{\omega}^\top\bx} (e^{-i \boldsymbol{\omega}^\top\bx^\prime})^\ast \right) \right\rbrack \\
	& = \Exp_{p(\varphi)} \left\lbrack  \varphi(\bx)^\top \varphi(\bx^\prime)  \right\rbrack .
	\label{fourier_trans1}
\end{align}
Here, the superscript $\ast$ denotes the complex conjugate and $\varphi(\cdot)$ is a random feature map. Note that the imaginary component is not required since we only work with real-valued kernels. A possible choice for $\varphi(\cdot)$ is
\begin{equation}
	\varphi(\bx) = \sqrt{2} \cos\left(\boldsymbol{\omega}^\top \bx + b\right) ,
	\label{rand_feature_map}
\end{equation}
in which $b \sim \mathcal{U} [0, 2\pi]$ is a uniform random variable \cite{rahimi2008, henrandez-lobato2014}. The distribution of $\boldsymbol{\omega}$ depends on the type of correlation function. For example, the spectral density $p(\boldsymbol{\omega})$ of the Mat\'ern kernel is a $t$-distribution and for the SE kernel is Gaussian specified by \cite{rahimi2008, wang2016}
\begin{equation}
	\boldsymbol{\omega}_{SE} \sim \mathcal{N} \left(\mathbf{0},  \boldsymbol{\Delta}^{-2} \right) .
	\label{omega_SE_kernel}
\end{equation}

The explicit random feature map $\varphi(\bx)$ defined by \Cref{rand_feature_map} allows us to estimate  the (actual) feature map $\boldsymbol\phi(\bx)$, which is possibly infinite dimensional. This can be performed using a Monte Carlo approach where we generate $M$ independently and identically distributed (i.i.d.) samples from $p(\boldsymbol{\omega})$ and $p(b) = \mathcal{U}[0, 2\pi]$ denoted by $\boldsymbol{\omega}^{(1)}, \ldots, \boldsymbol{\omega}^{(M)}$ and $b^{(1)}, \ldots, b^{(M)}$, respectively.  Then, the approximated feature map $\hat{\boldsymbol\phi}(\bx)$ is achieved by  
\begin{equation}
	\hat{\boldsymbol\phi}(\bx) = \sqrt{\frac{2}{M}} \left\lbrack \cos\left(\boldsymbol{\omega}^{(1) \top} \bx + b^{(1)} \right), \ldots, \cos\left(\boldsymbol{\omega}^{(M) \top} \bx + b^{(M)} \right)  \right\rbrack^\top ,
	\label{approx_phi}
\end{equation}
which transforms an input vector $\bx$ into the $M$-dimensional feature space. Finally, the stationary kernel $k$ is approximated as 
\begin{equation}
	k(\bx, \bx^\prime) = \Exp_{p(\varphi)} \left\lbrack  \varphi(\bx)^\top \varphi(\bx^\prime)  \right\rbrack \approx \hat{\boldsymbol\phi}(\bx)^\top \hat{\boldsymbol\phi}(\bx^\prime) .
\end{equation}
It is worth mentioning that the quality of the above approximation is 
\begin{equation}
	\sup_{\bx, \bx^\prime \in \mathcal{X}} \left| k(\bx, \bx^\prime) -  \hat{\boldsymbol\phi}(\bx)^\top \hat{\boldsymbol\phi}(\bx^\prime) \right| \le \varepsilon ,
	\label{kernel_approx_quality}
\end{equation} 
where $\varepsilon := \mathcal{O}(M^{-1/2})$ \cite{rahimi2008}.
%===========================================================================
\subsection{Generating GP sample paths} 
\label{sec:approx_pred_mean}
%===========================================================================
GPs can be interpreted from a \emph{weight space} perspective that is a weighted sum of (possibly infinite) basis functions. Under this view, the GP $Y(\bx)$ given in \Cref{stochastic_process} can be approximated as
\begin{equation}
	\hat{Y}(\bx) = \mu(\bx) + \hat{\boldsymbol\phi}(\bx)^\top \mathbf{w} ,
	\label{weight_view}
\end{equation}
with weights $\mathbf{w}\sim \mathcal{N}(\mathbf{0}, \sigma^2\mathbf{I})$ \cite{GPML}. The predictive distribution of $\hat{Y}(\bx)$ relies on the posterior of the weights, which is Gaussian characterised by 
\begin{align}
	\label{post_mean_weight}
	&m_\mathbf w = \Exp \left \lbrack \mathbf{w} \mid \mathcal{D} \right\rbrack = \boldsymbol{\Phi} \left(\boldsymbol{\Phi}^\top \boldsymbol{\Phi} \right)^{-1}  \left( \mathbf{y} -   \boldsymbol{\mu} \right) , \\
	\label{post_var_weight}
	&\Sigma_\mathbf w = \Cov \left( \mathbf{w} \mid \mathcal{D}  \right) = \sigma^2 \left( \mathbf{I} - \boldsymbol{\Phi} \left(\boldsymbol{\Phi}^\top \boldsymbol{\Phi} \right)^{-1} \boldsymbol{\Phi}^\top \right) .
\end{align}
In the above equations, $\boldsymbol{\Phi} = \left\lbrack\hat{\boldsymbol\phi}(\bx^1), \ldots, \hat{\boldsymbol\phi}(\bx^n)\right\rbrack$ is an $M\times n$-dimensional matrix and is the aggregation of columns of $\hat{\boldsymbol\phi}(\bx)$ for all points in the training set \cite{vanderwilk2019}. 

Let $\mathbf{w}^{(s)}$ be a realisation from $\mathcal{N}(m_\mathbf w, \Sigma_\mathbf w)$. A posterior sample from $\hat{Y}(\bx)$, denoted by $\hat{Y}^{(s)}(\bx)$, is given by
\begin{equation}
	\hat{Y}^{(s)}(\bx) = \mu(\bx) + \hat{\boldsymbol\phi}(\bx)^\top \mathbf{w}^{(s)} .
	\label{GP_post_sample}
\end{equation}
It is worth mentioning that generating $\hat{Y}^{(s)}(\bx)$ incurs a constant cost of $\mathcal O (M^3)$. However, sampling from a GP posterior distribution on a discrete domain $\mathcal X$ has a computational complexity of $\mathcal O(|\mathcal X|^3)$ due to a required Cholesky decomposition of the covariance matrix. The computational burden of this sampling strategy becomes prohibitive as $|\mathcal X|$ grows exponentially with the dimension \cite{mutny2018}.

The procedure to generate a sample of the GP posterior is outlined in Algorithm \ref{alg:approx_pred_mean}. Now, one can generate multiple such GP samples by drawing different realisations $\mathbf{w}^{(s)}$. This idea is used to emulate dynamical simulators where draws from the emulated flow map are employed to perform one-step ahead predictions. With this, we can quantify uncertainty of the time series prediction as described in the next section. 
%================================================
\begin{algorithm}[htpb]
	\caption{Drawing a posterior sample from a GP}
	\label{alg:approx_pred_mean}
	Input: training set $\lbrace \mathbf{X}, \mathbf{y} \rbrace$, trend function $\mu$, kernel spectral density $p(\boldsymbol{\omega})$, dimension of random feature space $M$
	\begin{algorithmic}[1]
		\STATE  Draw $M$ i.i.d. samples from $p(\boldsymbol{\omega})$ and $p(b) = \mathcal{U}\lbrack 0, 2\pi \rbrack$
		\STATE Construct $\hat{\boldsymbol\phi}(\bx)$ using (\Cref{approx_phi})
		\STATE Calculate $\hat{\boldsymbol\phi}(\bx)$  at the inputs to obtain $\boldsymbol{\Phi} = \left\lbrack \hat{\boldsymbol\phi}(\bx^1), \ldots, \hat{\boldsymbol\phi}(\bx^n) \right\rbrack$ 
		\STATE Draw the realisation $\mathbf{w}^{(s)}\sim \mathcal{N}(m_\mathbf w, \Sigma_\mathbf w)$ based on (\Cref{post_mean_weight}) and (\Cref{post_var_weight})
		\STATE Compute $\hat{Y}^{(s)}(\bx)$ using (\Cref{GP_post_sample})
	\end{algorithmic}
\end{algorithm}
%================================================
%===================================================================================================
\section{Emulating dynamical simulators}
\label{sec:Emul_Dynam_Sys}
%===================================================================================================
Now let us introduce our methodology for emulating deterministic nonlinear dynamical simulators that are based on ODEs. Let $\bx(t_1) = \left(x_1(t_1), \ldots, x_d(t_1) \right) ^\top$ be the solution of the system at $t_1 = t_0 + \Delta t$ for a given fixed ``small" time step $\Delta t$ and initial condition $\bx_0$. We assume that $\bx(t_1)$ is produced by the flow map $\textbf F$ defined as
\begin{equation}
	\textbf F(\bx_0) = \left(f_1(\bx_0), \ldots, f_d(\bx_0) \right)^\top = \left(x_1(t_1), \ldots, x_d(t_1) \right) ^\top ,
	\label{flow_map_component}
\end{equation}
such that each map $f_i: \mathcal{X} \mapsto \mathbb{R}$ yields the $i$-th component of $\bx(t_1)$: $f_i(\bx_0) = x_i(t_1)$. A prediction associated with the dynamics of $x_i(t)$ is achieved by:
\begin{itemize}
	\item Emulating $f_i(\bx)$ by a GP, i.e., $Y_i(\bx)$
	\item Using Algorithm \ref{alg:approx_pred_mean} to draw an approximate sample of $Y_i(\bx)$, i.e., $\hat{Y}^{(s)}_i(\bx)$ 
	\item Using $\hat{Y}^{(s)}_i(\bx)$ iteratively to perform one-step ahead predictions
\end{itemize}

\begin{figure}[htpb] 
	\begin{minipage}{0.45\textwidth}
		\includegraphics[width=0.98\textwidth]{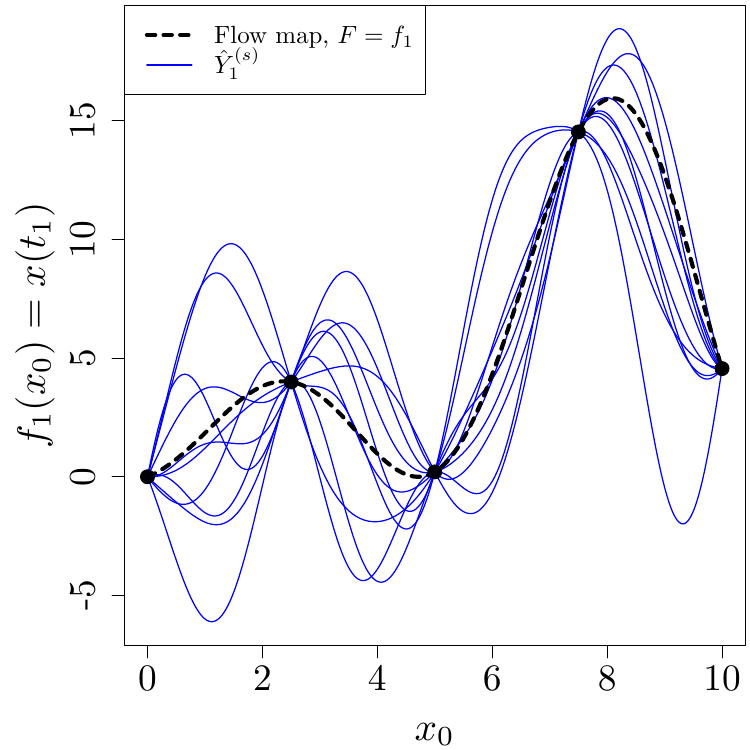}
	\end{minipage}
	\begin{minipage}{0.54\textwidth}
		\begin{tikzpicture}[scale=1, transform shape]
			\node (x0) {$x_0$};
			\node[right=1cm of x0] (x1) {$\hat{x}^{(1)}(t_1)$};
			\path[->]
			(x0) edge[bend left=45] node[pos=0.5,above] {$\hat{Y}^{(1)}_1$} (x1);
			\node[right=1cm of x1] (x2) {$\hat{x}^{(1)}(t_2)$};
			\path[->]
			(x1) edge[bend left=45] node[pos=0.5,above] {$\hat{Y}^{(1)}_1$} (x2);
			\node[right=1cm of x2] (x3) {$\hat{x}^{(1)}(t_3)$};
			\path[->]
			(x2) edge[bend left=45] node[pos=0.5,above] {$\hat{Y}^{(1)}_1$} (x3);
			\node[right=0.1cm of x3] (x4) {$\ldots$};
		\end{tikzpicture}	
		
		\begin{tikzpicture}[scale=1, transform shape]
			\node (x0) {$x_0$};
			\node[right=1cm of x0] (x1) {$\hat{x}^{(2)}(t_1)$};
			\path[->]
			(x0) edge[bend left=45] node[pos=0.5,above] {$\hat{Y}^{(2)}_1$} (x1);
			\node[right=1cm of x1] (x2) {$\hat{x}^{(2)}(t_2)$};
			\path[->]
			(x1) edge[bend left=45] node[pos=0.5,above] {$\hat{Y}^{(2)}_1$} (x2);
			\node[right=1cm of x2] (x3) {$\hat{x}^{(2)}(t_3)$};
			\path[->]
			(x2) edge[bend left=45] node[pos=0.5,above] {$\hat{Y}^{(2)}_1$} (x3);
			\node[right=0.1cm of x3] (x4) {$\ldots$};
			\node[below=0.4cm of x1] {$\vdots$};
			\node[below=0.4cm of x2] {$\vdots$};
			\node[below=0.4cm of x3] {$\vdots$};
		\end{tikzpicture}	
		
		\begin{tikzpicture}[scale=1, transform shape]
			\node (x0) {$x_0$};
			\node[right=1cm of x0] (x1) {$\hat{x}^{(S)}(t_1)$};
			\path[->]
			(x0) edge[bend left=45] node[pos=0.5,above] {$\hat{Y}^{(S)}_1$} (x1);
			\node[right=1cm of x1] (x2) {$\hat{x}^{(S)}(t_2)$};
			\path[->]
			(x1) edge[bend left=45] node[pos=0.5,above] {$\hat{Y}^{(S)}_1$} (x2);
			\node[right=1cm of x2] (x3) {$\hat{x}^{(S)}(t_3)$};
			\path[->]
			(x2) edge[bend left=45] node[pos=0.5,above] {$\hat{Y}^{(S)}_1$} (x3);
			\node[right=0.1cm of x3] (x4) {$\ldots$};
		\end{tikzpicture}	
	\end{minipage}
	\caption{Left: The flow map function $F(x_0) = f_1(x_0) = x(t_1)$ (dashed) and approximate GP sample paths (blue) obtained by RFF. Right: One-step ahead predictions using sample paths $\hat{Y}^{(s)}_1, s = 1, \ldots, S,$ in an iterative manner.}
	\label{fig:1Dexample}
\end{figure}

Following the above procedure renders only one prediction of the time series. However, we wish to have an estimation of uncertainty associated with the prediction accuracy. This can be achieved by repeating the above steps with different draws from the emulated flow map to obtain a distribution over the time series. The mean and variance of that distribution at a given time point serve as the model output prediction and the associated uncertainty there, respectively. More rigorously, let $\hat{x}_i(t)$ be the model output prediction corresponding to the $i$-th component of $\textbf F$, and $SD\left(\hat{x}_i(t)\right)$ represent the corresponding standard deviation at time $t= t_1, \ldots, T$, with $T$ being the final time of the simulation. We can write
	\begin{equation}
		\hat{x}_i(t) = \frac{\sum_{s = 1}^{S} \hat{x}^{(s)}_i(t)}{S}, \quad SD\left(\hat{x}_i(t)\right) = \sqrt{\frac{\sum_{s = 1}^{S} \left( \hat{x}^{(s)}_i(t) - \hat{x}_i(t) \right)^2 }{S - 1}} ,
		\label{time_series_pred}
	\end{equation}
	in which $S$ represents the number of sample paths taken from the emulated flow map, and $\hat{x}^{(s)}_i(t)$ is the prediction obtained via $\hat{Y}^{(s)}_i(\bx)$. To shed more light on the proposed approach, an illustrative example is provided in \Cref{fig:1Dexample}. The pictures show how the approximate GP samples (blue) obtained by RFF are used to perform one-step ahead predictions.

%###############################################################################
\begin{algorithm}[!hb] 
	\caption{Emulating dynamic nonlinear simulators} \label{alg:emulation_method}
	\begin{algorithmic}[1]
		\STATE Select $n$ ``space-filling" sample initial conditions: $\mathbf{X} = \left\lbrace \mathbf{x} ^{1}_0, \dots , \mathbf{x} ^{n}_0 \right\rbrace $  
		\STATE Run the simulator for each $\mathbf{x} ^{1}_0, \dots , \mathbf{x} ^{n}_0$ over $\Delta t$ to obtain\newline $\mathbf{y} = \left\lbrace \mathbf{x} ^{1}(t_1), \dots , \mathbf{x} ^{n}(t_1) \right\rbrace$
		\FOR {$i=1$ to $d$}
		\STATE Create the training set $\mathcal{D}_i = \lbrace \mathbf{X}, \mathbf{y}_i \rbrace$ where $\mathbf{y}_i = \left( x_i ^{1}(t_1), \dots , x_i ^{n}(t_1) \right)^\top$
		\STATE Build the emulator $Y_i(\bx)$ for $f_i(\cdot)$, the $i$-th component of $\textbf F$
		\FOR {$s=1$ to $S$} 
		\STATE Generate the approximate sample path $\hat{Y}_i^{(s)}(\bx)$ using Algorithm \ref{alg:approx_pred_mean} 
		\STATE Use $\hat{Y}_i^{(s)}(\bx)$ iteratively to perform one-step ahead predictions
		\ENDFOR
		\STATE Use (\Cref{time_series_pred}) to calculate the prediction and associated uncertainty for $x_i(t)$,  $t = t_1, \ldots, T$
		\ENDFOR
	\end{algorithmic}
\end{algorithm}
%############################################################################### 

Our proposed emulation method is summarised in Algorithm \ref{alg:emulation_method}. The first step in Algorithm \ref{alg:emulation_method} is to choose $n$ training points, $\bx_0^1, \ldots, \bx_0^n$, from the space of initial condition $\mathcal{X}$. The accuracy of emulators depends substantially on the location of these points, and their corresponding outputs obtained in the next step. They are used to train GPs as described in \Cref{sec:GPR} where the unknown parameters are estimated via maximising the likelihood function. In this work, the sampled initial conditions are selected in a carefully designed experiment so that they fill the space as uniformly as possible. This sampling strategy is called a ``space-filling" design, where we use the \texttt{maximinSA\_LHS} function implemented in the R package \texttt{DiceDesign} \cite{DiceDesign}. The \texttt{maximinSA\_LHS} function produces such designs by maximising the minimum distance between all candidate points. Interested readers are referred to \cite{sacks1989, santner2003} for more information on space-filling designs.
%===================================================================================================
\section{Numerical results}
\label{sec:Num_Res}
%===================================================================================================
The prediction performance of our proposed method is tested on several dynamical systems implemented as computer codes. They are the Lorenz \cite{lorenz1963}, van der Pol \cite{strogatz2007}, and Hindmarsh-Rose models \cite{hindmarsh1984} and are further elaborated in the following subsections. The training data set consists of $n = 15\times d$ space-filling points selected from the space of initial conditions. Generally, increasing the size of the training set results in a more accurate emulation. However, in the context of computationally expensive simulators, we have access to a limited number of model runs. We use a training set of size $n = 15\times d$, which appears to be adequate according to our experiments, suggesting it as a recommended rule of thumb.

The GP correlation function is the squared exponential kernel as is recommended in \cite{conti2009, mohammadi2019}. The trend function is a first order regression model: $\mu(\bx_0) = q(\bx_0)^\top\boldsymbol{\beta}$ with $q(\bx_0) = \lbrack 1, \bx_0^\top \rbrack$. The regression coefficients ($\boldsymbol{\beta}$) together with the covariance function parameters such as the length-scales ($\boldsymbol{\delta}$) and process variance ($\sigma^2$) are estimated using the maximum likelihood method, see \Cref{sec:GPR}. These parameters are separately estimated for each element of the state vector using data from the initial time step as elaborated in the previous section. The number of random features ($M$) determines the quality of Monte Carlo approximation of the kernel, see \Cref{kernel_approx_quality}. In \cite{rahimi2008a, liu_kernel2022} it is shown that using a number of random features proportional to the size of the training set ($\Omega(n)$) is sufficient to achieve a comparable performance to that of the original kernel. We set $M = 250$ which is higher than the size of the training sets in all our experiments and is also used in \cite{henrandez-lobato2014}. The number of realisations drawn from the approximate flow map is $S = 100$. Notice that $S$ is the number of simulated time series, and gives, at any time $t$, $S$ samples whose mean ($\hat x_i(t)$) serves as the model output prediction. In the specific scenario of sampling from a normal population, the distribution of $\hat x_i(t)$, with an unknown population variance, conforms to a Student's t-distribution with $S - 1$ degrees of freedom. The simulation time step is fixed and equal to $\Delta t = 0.01$. The ODE is solved on $\lbrack t_0, t_1\rbrack$ by the default solver of the R package \texttt{deSolve} \cite{soetaert2010}.

The accuracy of the time series prediction is measured via the \emph{mean absolute error (MAE)} and \emph{root mean square error (RMSE)} criteria. They are defined as
\begin{align}
	&MAE = \frac{\sum_{t = 1}^{t = T}  \left\lvert x_i(t) - \hat{x}_i(t) \right\rvert}{n_{step}} , \\
	&RMSE = \sqrt{\frac{\sum_{t = 1}^{t = T} \left( x_i(t) - \hat{x}_i(t) \right)^2}{n_{step}}} ,
\end{align}
for $i = 1, \ldots, d$. Here, $n_{step} = T/\Delta t$ denotes the total number of one-step ahead predictions. We compare our results with the method presented in \cite{mohammadi2019} as it is also based on emulating the numerical flow map and one-step ahead predictions. In this method, each component of the flow map function ($f_i, i = 1, \ldots, d$) is emulated by the standard GP paradigm over the initial time step $\Delta t = t_1 - t_0 $. The selection of training data follows a similar approach to the method proposed in this paper. The unknown parameters of the emulator are estimated using maximum likelihood estimation and therefore the method is not fully Bayesian. Let $\hat{f}_i$ be the GP emulator associated with $f_i$. The prediction of $x_i(t)$ relies on the iterative use of $\hat{f}_i$, following the Markov assumption, for one-step ahead predictions across the time horizon $T$. Given that GP prediction results in a distribution (as outlined in Equations (\ref{GP_mean}) and (\ref{GP_var})), in this approach, only the initial input to $\hat{f}_i$ is certain. From $t_1$ onwards, the input to the emulator is actually the output from the previous step. To account for the input uncertainty, at each iteration, samples are repeatedly drawn from the uncertain inputs using the Monte Carlo method. These samples are then propagated through the emulator, and the output distribution is approximated as a normal distribution using the law of iterated expectations and conditional variance. We refer to our proposed approach as ``Method 1" and that of \cite{mohammadi2019} as ``Method 2" in the rest of the paper. 
%=======================================================================
\subsection{Lorenz system} 
\label{sec:lorenz}
%=======================================================================
The Lorenz system was initially derived by Edward Lorenz in 1963 \cite{lorenz1963}. Although it was originally developed as a model of convection in the earth's atmosphere, the Lorenz system has applications to other fields, see e.g. \cite{peters1991}. This model can produce a famous chaotic attractor whereby trajectories on the attractor are very sensitive to initial conditions. In other words, the difference in evolution of the system starting from two slightly different points will become large. The Lorenz equations are
%###############################################################################
\begin{equation}
	\begin{cases}
		\frac{d x_1}{d t} = a_1 x_1 + x_2x_3 \\  \frac{d x_2}{d t} = a_2(x_2 - x_3) \\  \frac{d x_3}{d t} = -x_1x_2 + a_3 x_2 - x_3  ,
	\end{cases}
	\label{lorenz}
\end{equation}
%###############################################################################
with the classic parameter values of $a_1 = -8/3, a_2 = -10$ and $a_3 = 28$. which are used in our experiments. These values result in the well-know ``butterfly attractor" \cite{viana2000}.

\begin{figure}[htpb] 
	\includegraphics[width=0.46\textwidth]{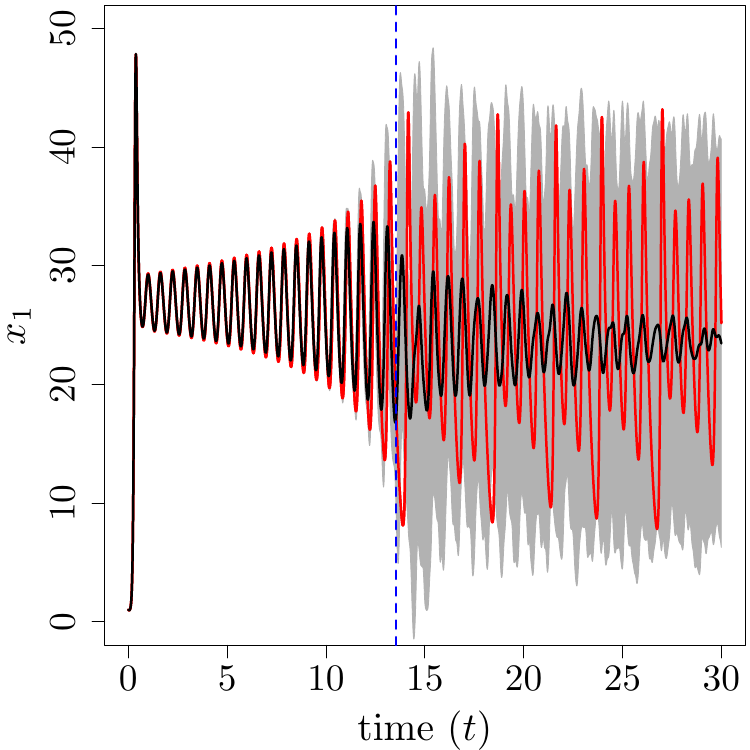}
	\includegraphics[width=0.46\textwidth]{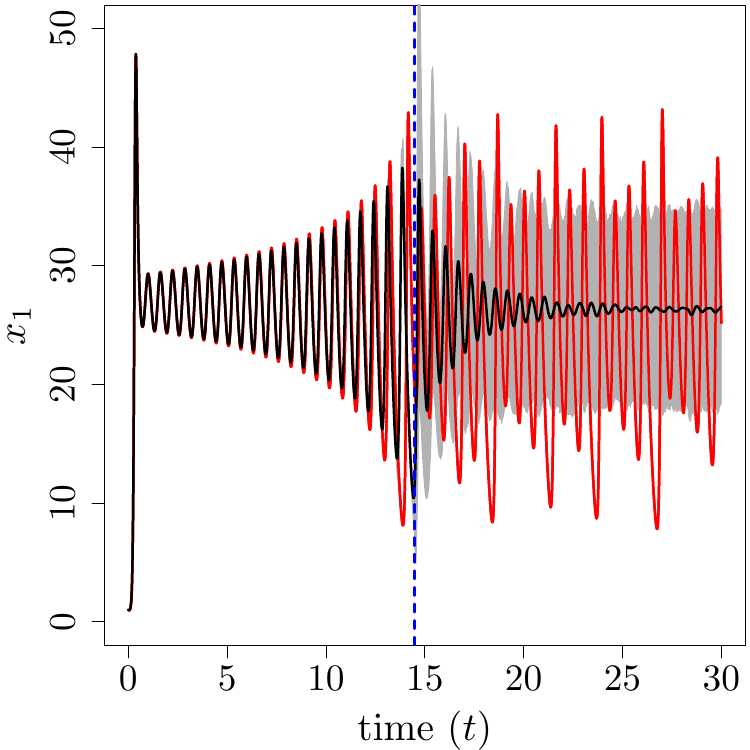}\\
	\includegraphics[width=0.46\textwidth]{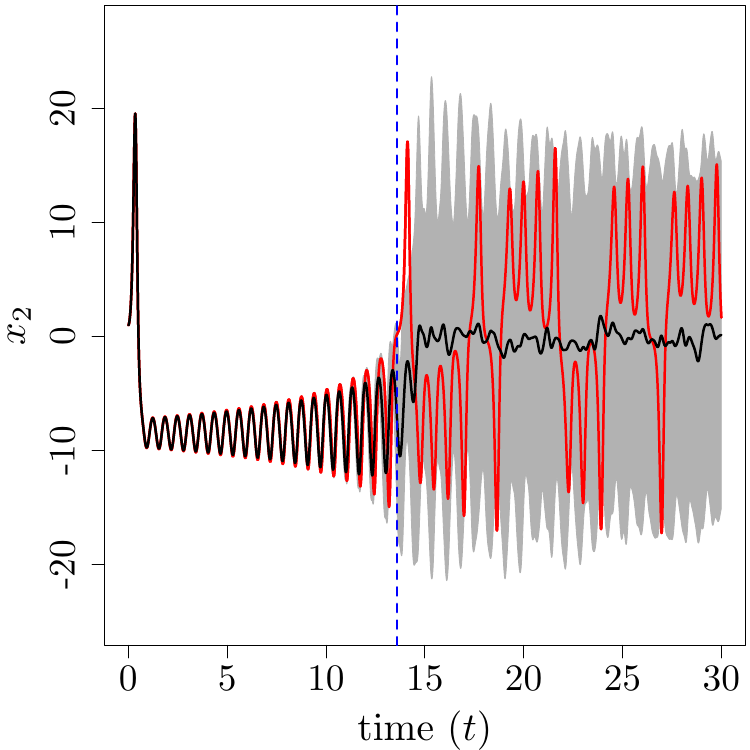}
	\includegraphics[width=0.46\textwidth]{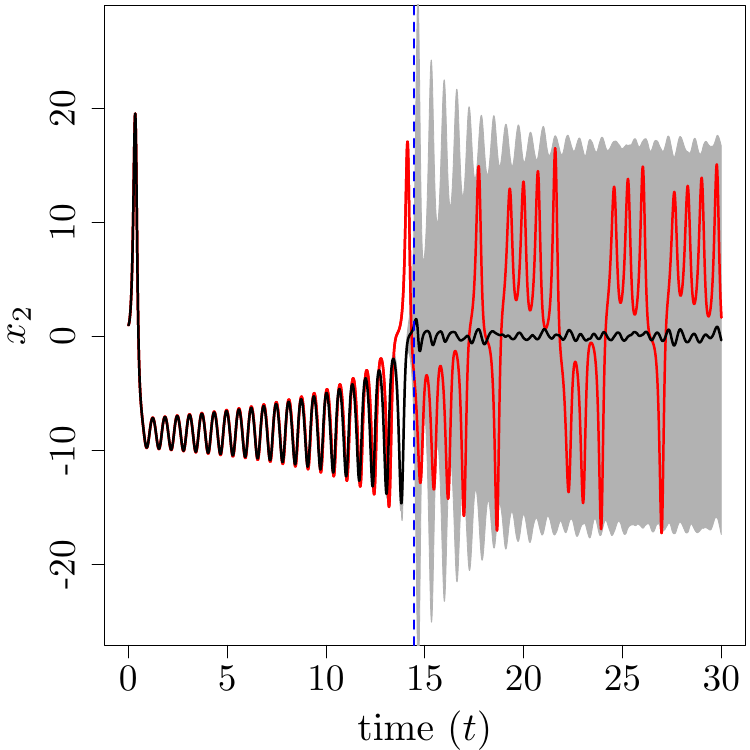}\\
	\includegraphics[width=0.46\textwidth]{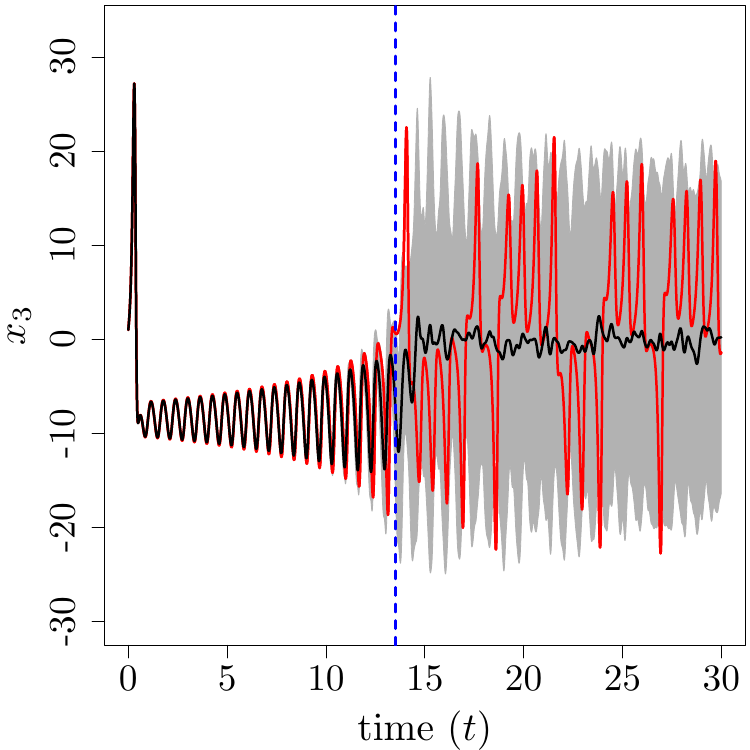}
	\includegraphics[width=0.46\textwidth]{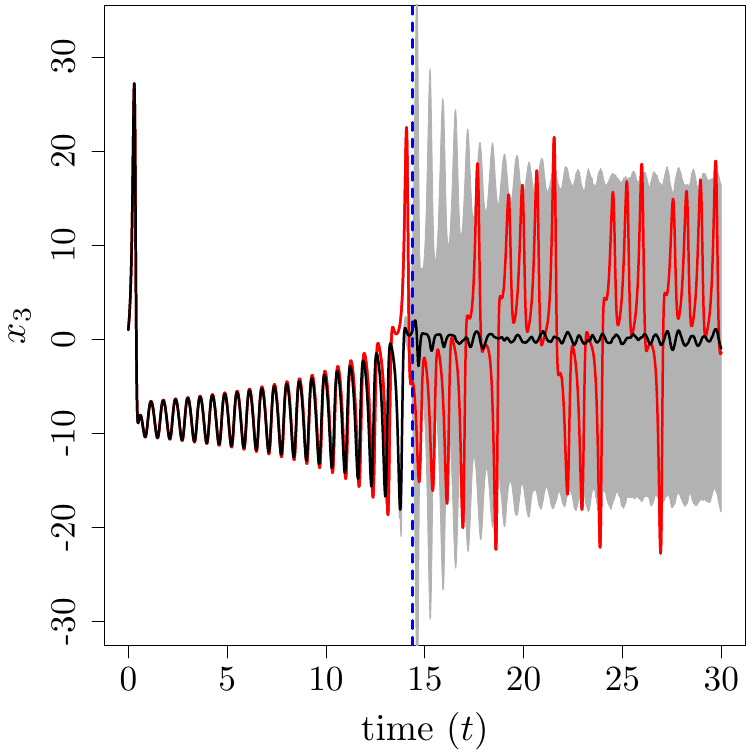}
	\caption{The prediction (black) and associated uncertainty (shaded) in emulating the Lorenz model (red) using Method 1 (left) and Method 2 (right). The initial condition is $\bx_0 = (1, 1, 1)^\top$. The vertical dashed blue lines represent the predictability horizon, and are the change point in the diagram of prediction uncertainties. The parameter values are $a_1 = -8/3, a_2 = -10$ and $a_3 = 28$.}
	\label{fig:lorenz}
\end{figure}

\begin{figure}[htpb] 
	\includegraphics[width=0.48\textwidth]{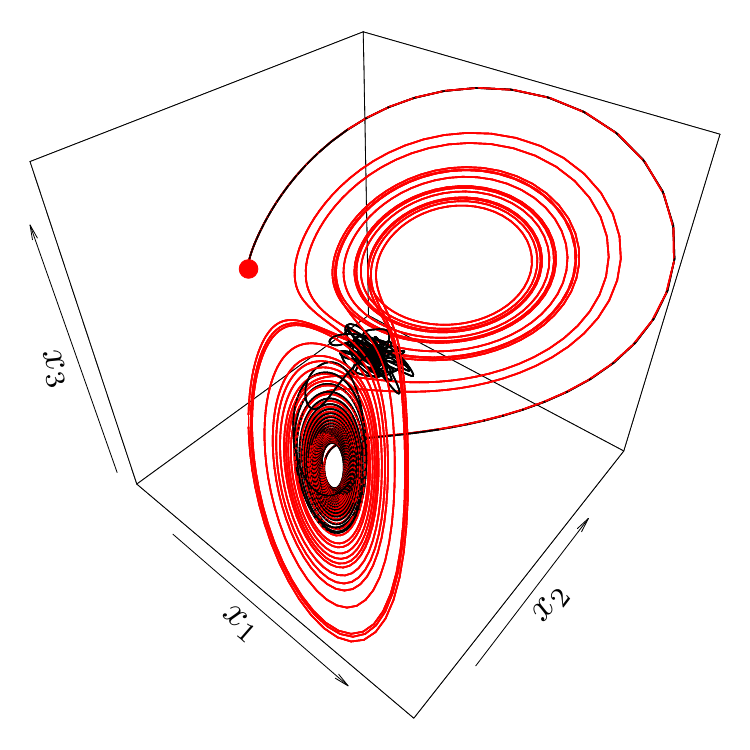}
	\includegraphics[width=0.48\textwidth]{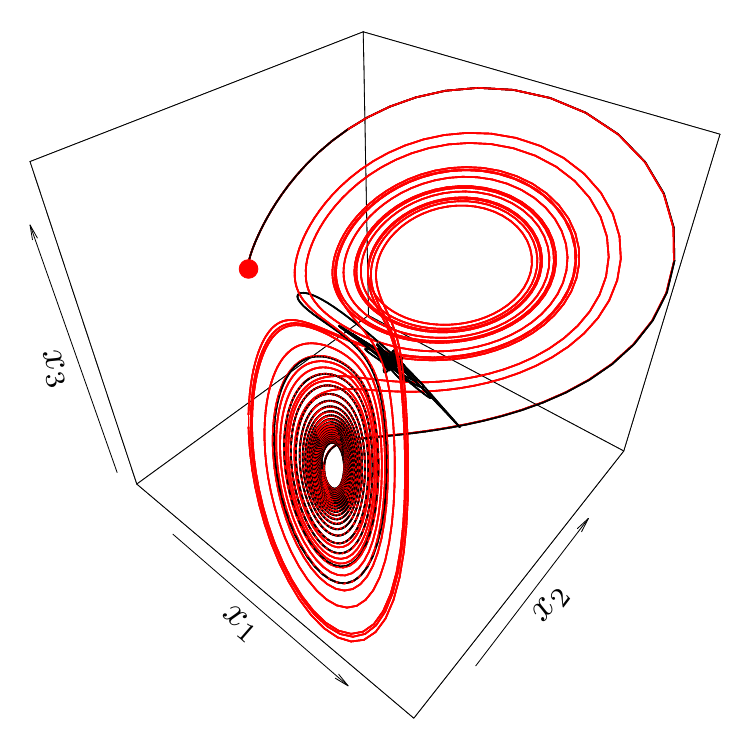}
	\caption{The three-dimensional graph of the Lorenz model (red) and its emulation (black) obtained by Method 1 (left) and Method 2 (right) following \Cref{fig:lorenz}. The red point indicates the initial condition that is $\bx_0 = (1, 1, 1)^\top$.}
	\label{fig:lorenz_3D}
\end{figure}

The predictive performance of Methods 1 (left) and 2 (right) in emulating the Lorenz model is displayed in Figures \ref{fig:lorenz} and \ref{fig:lorenz_3D}. The latter shows the corresponding three-dimensional graph where the red point is the initial condition, i.e., $\bx_0 = (1, 1, 1)^\top$. Throughout this paper, the simulation is shown in red, emulation in black and the shaded area represents the $95\%$ confidence interval: the prediction $\pm1.96 \times$ standard deviation (SD) of prediction. The two methods exhibit similar performances and their prediction accuracy is high up to around $t \approx 14$. After that time, the emulator deviates from the true model and tends to the average of the process. At the same time, the prediction uncertainties blow up, which allows the credible interval to encompass most values of the system. 

\begin{figure}[htpb] 
	\centering
	\includegraphics[width=0.53\textwidth]{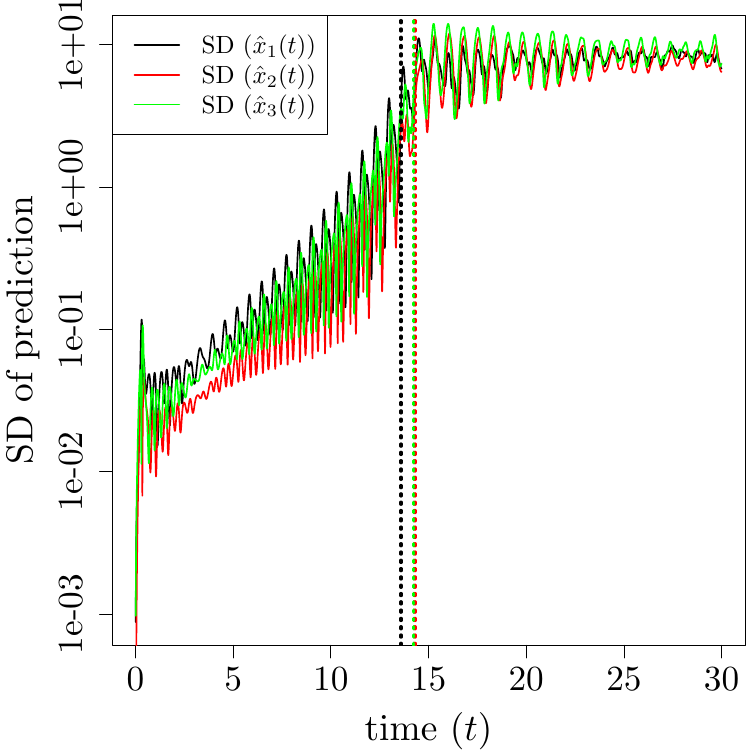}
	\caption{Standard deviation (SD) of prediction associated with the variables $x_1$ (black), $x_2$ (red), and $x_3$ (green) obtained by Method 1 in the Lorenz model. The vertical dotted lines show the change point of each diagram. The y-axis is on a logarithmic scale.}
	\label{fig:sd_predict}
\end{figure}

We note that the Lorenz attractor cannot be predicted perfectly due to its chaotic behaviour. The vertical dashed blue lines indicate the ``predictability horizon" defined as the time at which a change point occurs in the SD of prediction \cite{mohammadi2019}. The predictability horizon is acquired by applying the \texttt{cpt.mean} function implemented in the R package \texttt{changepoint} \cite{killick2012, killick2014} to the SD of predictions. This is depicted in \Cref{fig:sd_predict}, illustrating the SD of predictions for the state variables $x_1$ (black), $x_2$ (red), and $x_3$ (green) obtained by Method 1 in the Lorenz system. The vertical dotted lines are the corresponding change points. The predictability horizon looks slightly better for Method 2. However, Method 1 appears to capture the uncertainty better in that it encompasses all the trajectory and more tightly follows the fluctuation in dynamics. This can be investigated using the \emph{coverage probability} defined as the percentage of the times that the true model is within the $95\%$ uncertainty bounds. The coverage probability obtained by the two methods is given below.

\begin{center}
	\begin{tabular}{l || l c r}
		& $x_1$ & $x_2$ & $x_3$\\ 
		\hline
		Method 1 & 70.5 & 72.4 & 71.7 \\
		\hline
		Method 2 & 32.9 & 55.3 & 51.8
	\end{tabular}
	\newline
\end{center}  
%###############################################################################
%###############################################################################
%###############################################################################
%###############################################################################

\begin{figure}[htpb] 
	\includegraphics[width=0.48\textwidth]{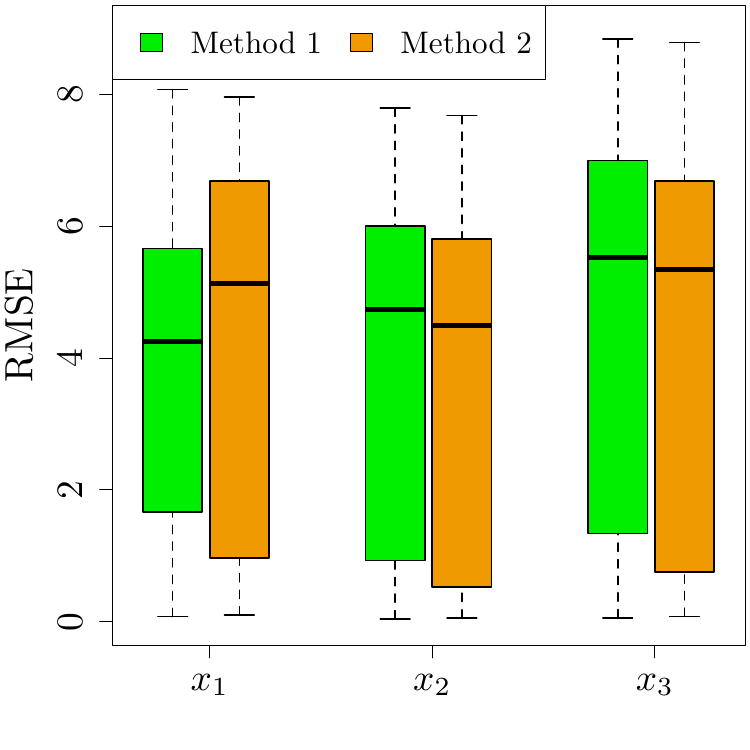}
	\hspace{0.1cm}
	\includegraphics[width=0.48\textwidth]{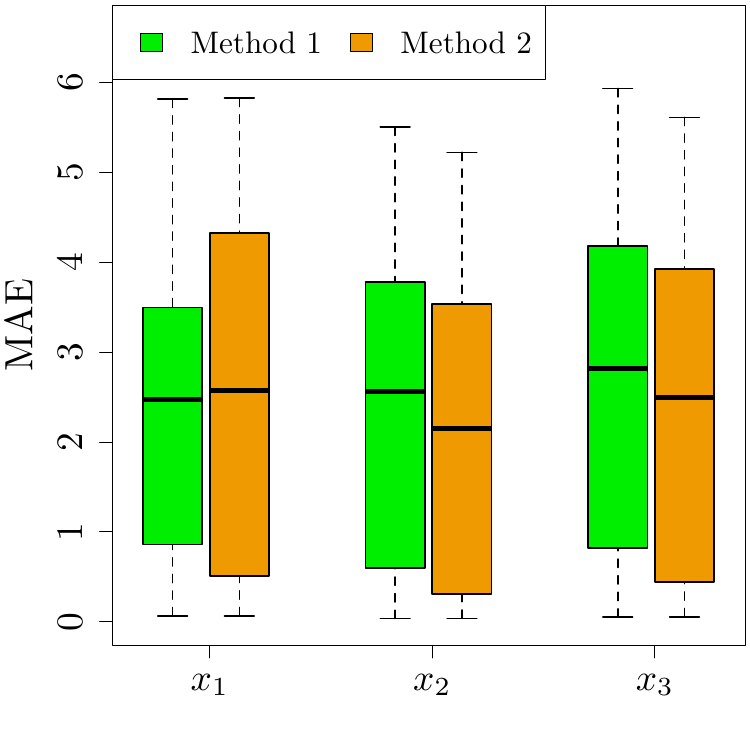}\\
	\caption{The box plot of the RMSE and MAE criteria in emulating the Lorenz system. The results are achieved using 100 initial conditions selected randomly from $\lbrack -10, 10\rbrack^3$. The accuracy of Method 1 (green) is higher than Method 2 (orange) in emulating $x_1$. However, they preform comparably when predicting $x_2$ and $x_3$.}
	\label{fig:lorenz_RMSE_MAE}
\end{figure}

\begin{figure}[htpb] 
	\includegraphics[width=0.32\textwidth]{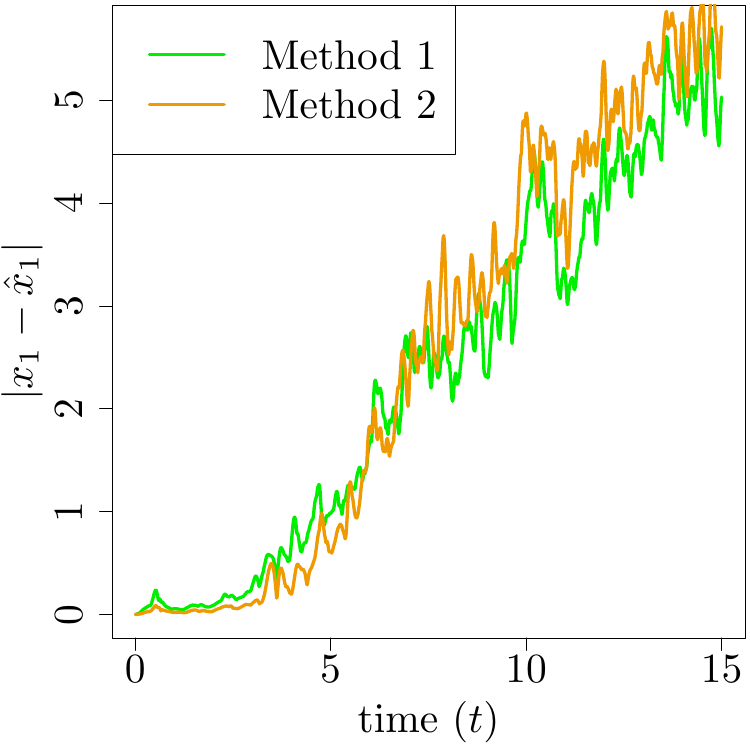}
	\includegraphics[width=0.32\textwidth]{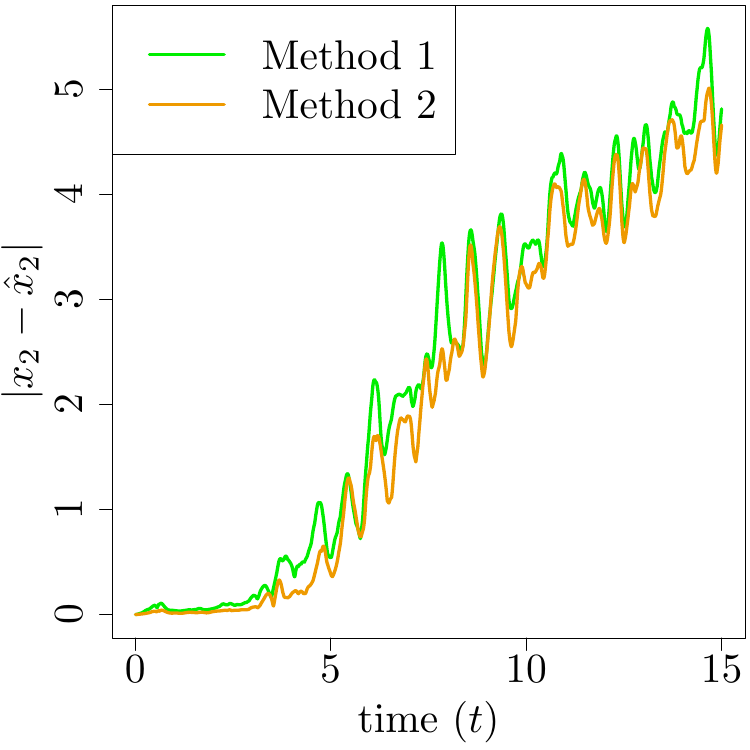}
	\includegraphics[width=0.32\textwidth]{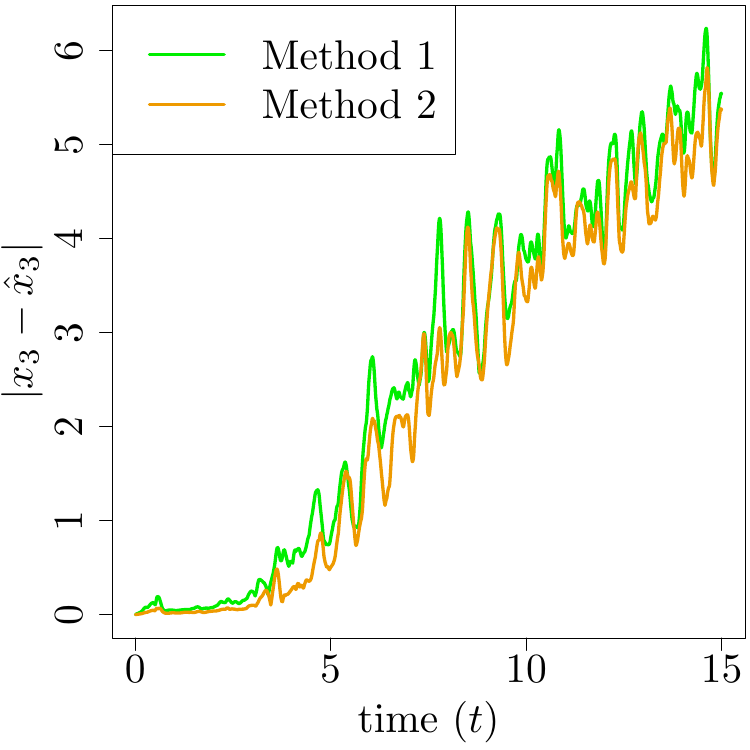}
	\caption{Accuracy trends of Methods 1 (green) and 2 (orange) over time for $x_1$ (left), $x_2$ (middle), and $x_3$ (right). Each curve represents the mean of $\lvert x_i - \hat{x}_i \rvert$ resulting from 100 initial conditions. The  vertical axis is on a logarithmic scale.}
	\label{fig:Lorenz_Accuracy_time}
\end{figure}

In order to have a more rigorous comparison between the two methods, we compare them based on the RMSE and MAE criteria. To do this, we select randomly 100 initial conditions from $\lbrack -10, 10\rbrack^3$ and for each initial condition we emulate the model with the two methods. \Cref{fig:lorenz_RMSE_MAE} illustrates the results obtained by Methods 1 (green) and 2 (orange). As can be seen, Method 1 has a higher prediction accuracy in emulating the first state variable ($x_1$) than Method 2. For the second and the third state variables, both methods have similar predictive performances.  \Cref{fig:Lorenz_Accuracy_time} presents how the accuracy of the two methods evolves over time. Each curve in the plot represents the mean of $\lvert x_i - \hat{x}_i \rvert, i = 1, \ldots, d,$ computed from 100 initial conditions. Both methods exhibit similar trends, however, Method 1 exhibits a better accuracy trend for $x_1$.
%###############################################################################   

%=======================================================================
\subsection{Van der Pol oscillator} 
\label{sec:vander}
%=======================================================================
The van der Pol equation was first proposed by the Dutch engineer Balthasar van der Pol in 1920 while working on the behaviour of vacuum tube circuits. Since then, this model has been extensively used to study oscillatory phenomena such as the heartbeat \cite{dossantos2004}, biological \cite{winfree1967} and circadian rhythms \cite{camacho2004}. The evolution of the van der Pol model are \cite{strogatz2007}
%###############################################################################
\begin{equation}
	\begin{cases}
		\frac{d x_1}{d t} = x_2\\  \frac{d x_2}{d t} = a(1 - x_1^2)x_2 - x_1 ,
	\end{cases}
	\label{van_der_pol}
\end{equation}
%###############################################################################
where the parameter $a > 0$ controls the frequency of oscillations. The van der Pol oscillator exhibits a periodic motion, i.e., a limit cycle. 

\begin{figure}[htpb] 
	\includegraphics[width=0.46\textwidth]{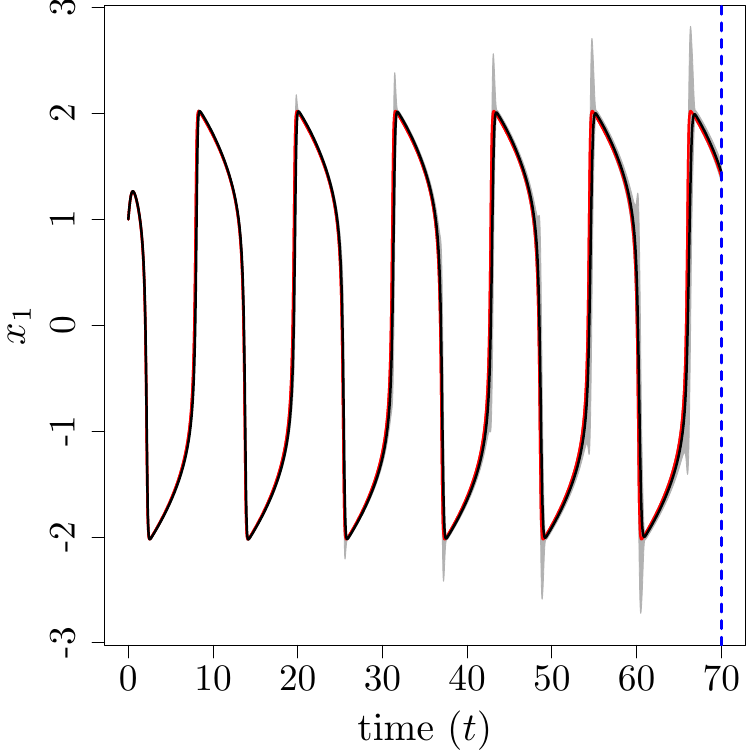}
	\includegraphics[width=0.46\textwidth]{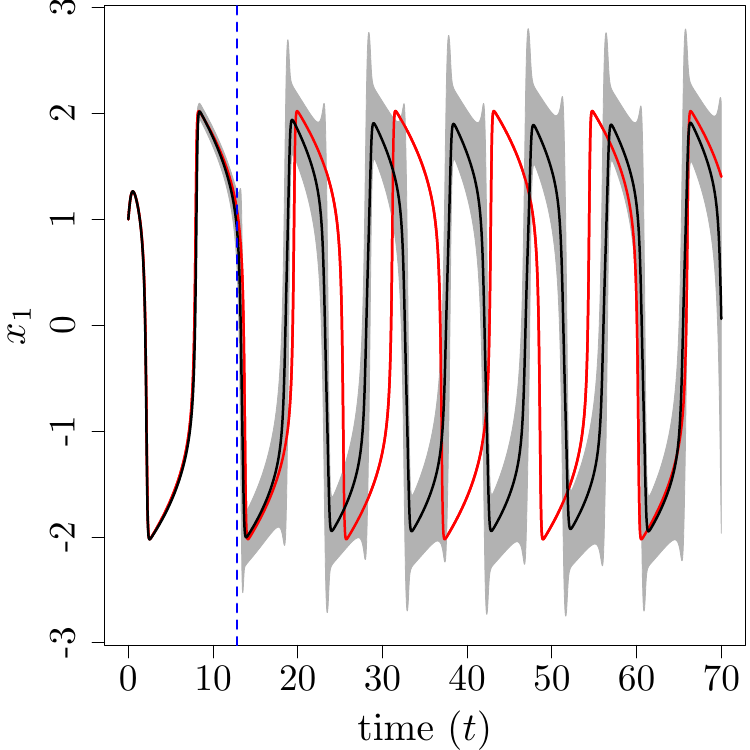}\\
	\includegraphics[width=0.46\textwidth]{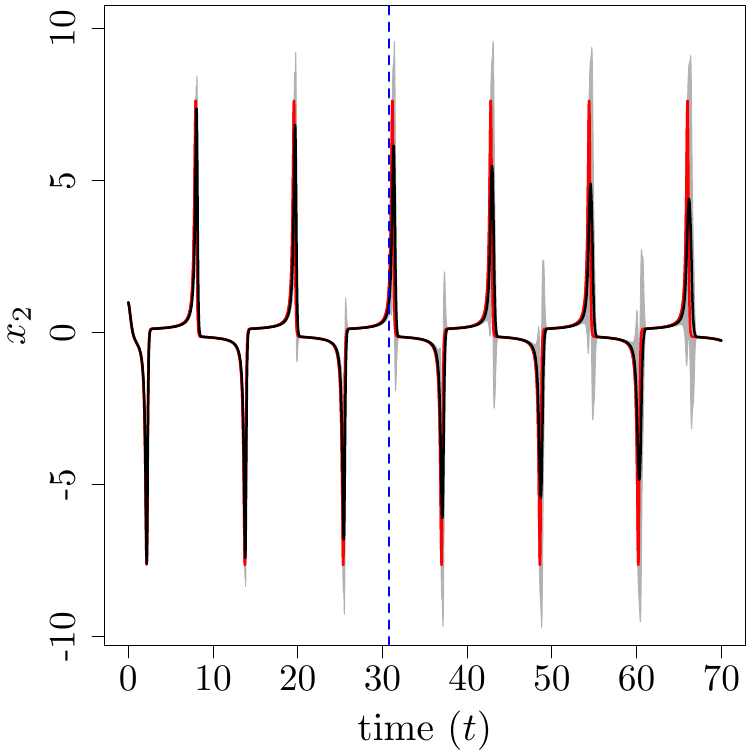}
	\includegraphics[width=0.46\textwidth]{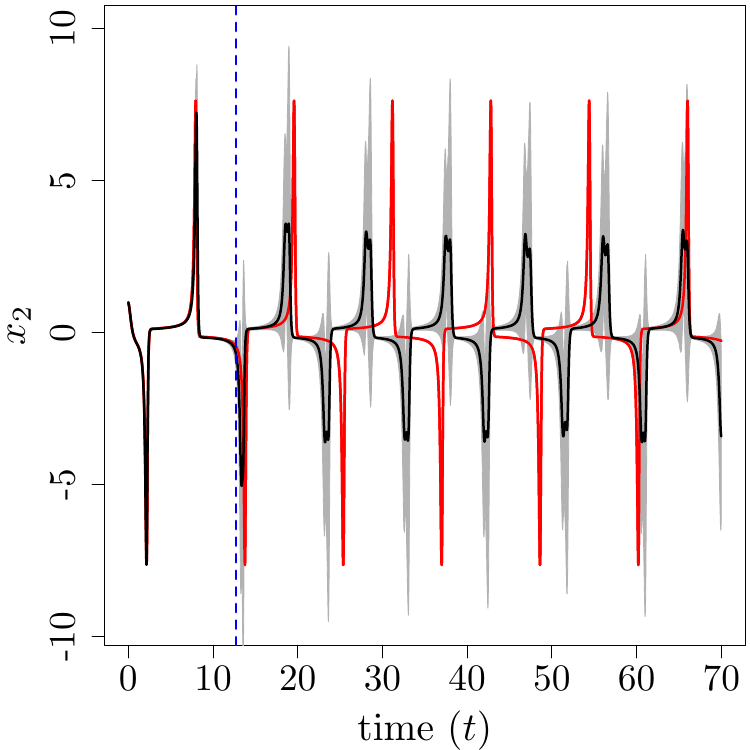}\\
	\includegraphics[width=0.46\textwidth]{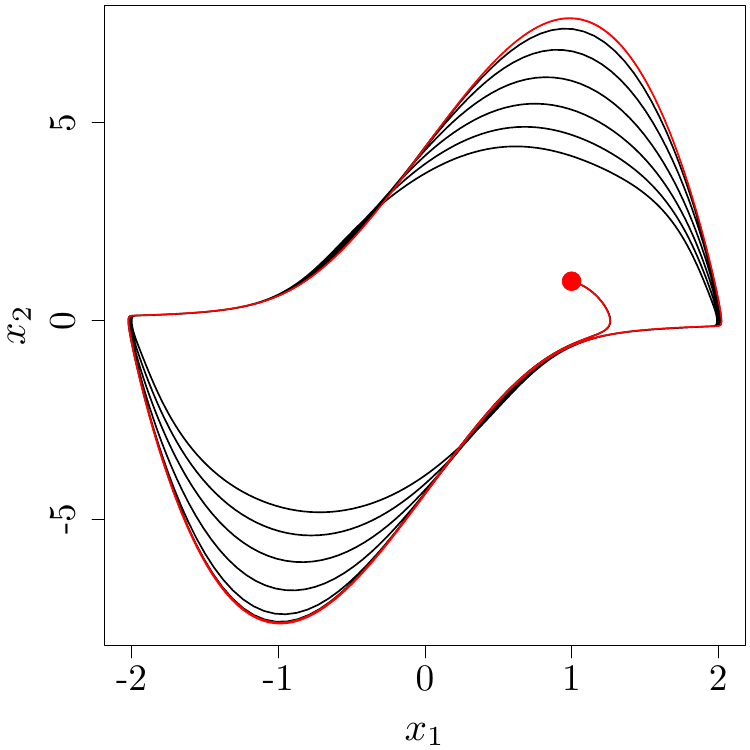}
	\includegraphics[width=0.46\textwidth]{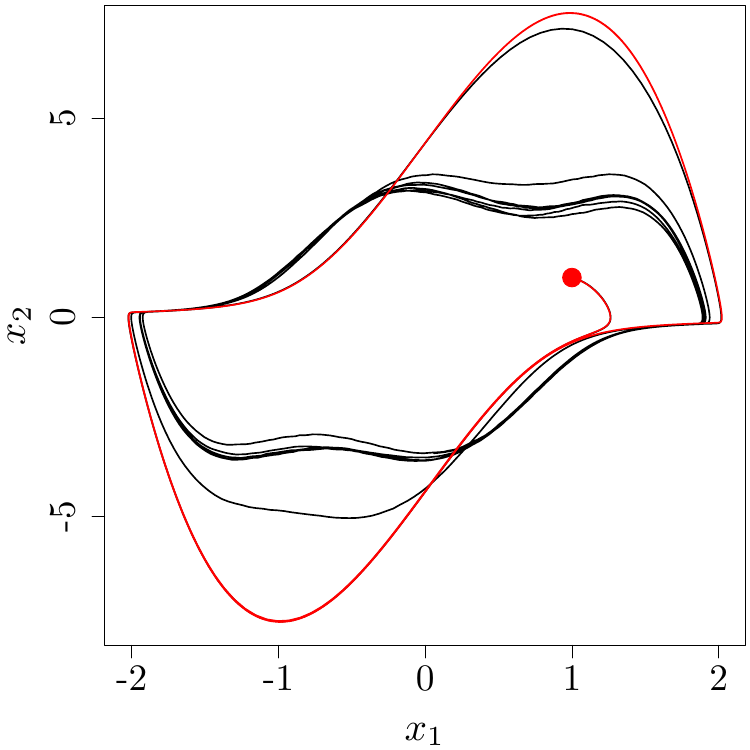}
	\caption{The van der Pol oscillator (red), the prediction (black) and credible interval (shaded) using Method 1 (left) and Method 2 (right). The dashed blue lines show the predictability horizon. The initial condition and parameter value are $\bx_0 = (1, 1)^\top$ and $a = 5$. The proposed approach in this paper outperforms Method 2 in emulating the van der Pol model.}
	\label{fig:vander}
\end{figure}

The emulation of the van der Pol oscillator based on Methods 1 (left) and 2 (right) is visualised in \Cref{fig:vander}. The truth (red), prediction (black), credible interval (shaded) and predictability horizon (dashed blue) are illustrated in the figures. The initial condition (the red point in the two-dimensional picture) is $\bx_0 = (1, 1)^\top$ and $a = 5$ in both cases. We observe that Method 1 outperforms Method 2 in emulating the van der Pol model. In particular, the first state variable is predicted with a high accuracy by Method 1 and the predictability horizon is equal to the total simulation time. Also, the coverage probability obtained by Method 1 is significantly better than Method 2 as computed below.

\begin{figure}[htpb] 
	\includegraphics[width=0.48\textwidth]{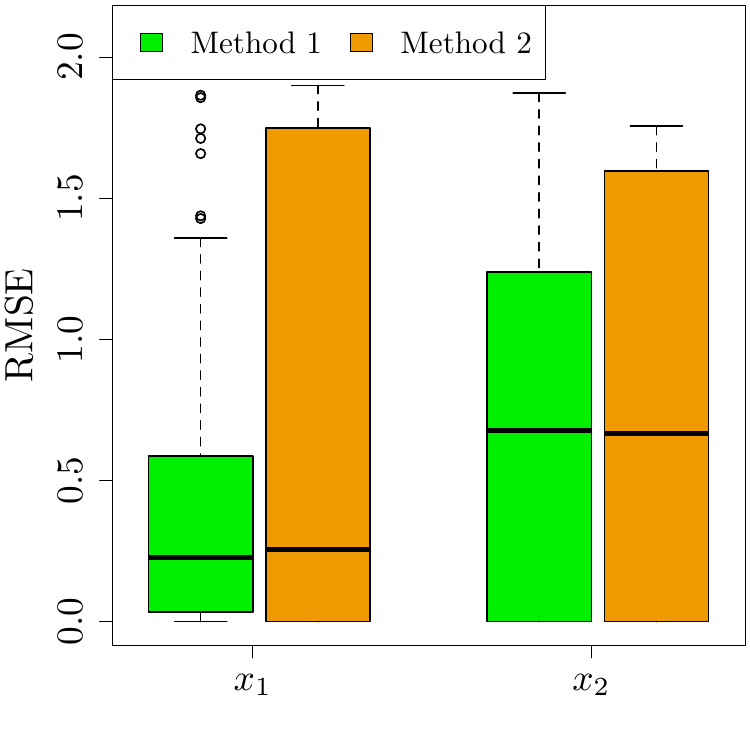}
	\hspace{0.2cm}
	\includegraphics[width=0.48\textwidth]{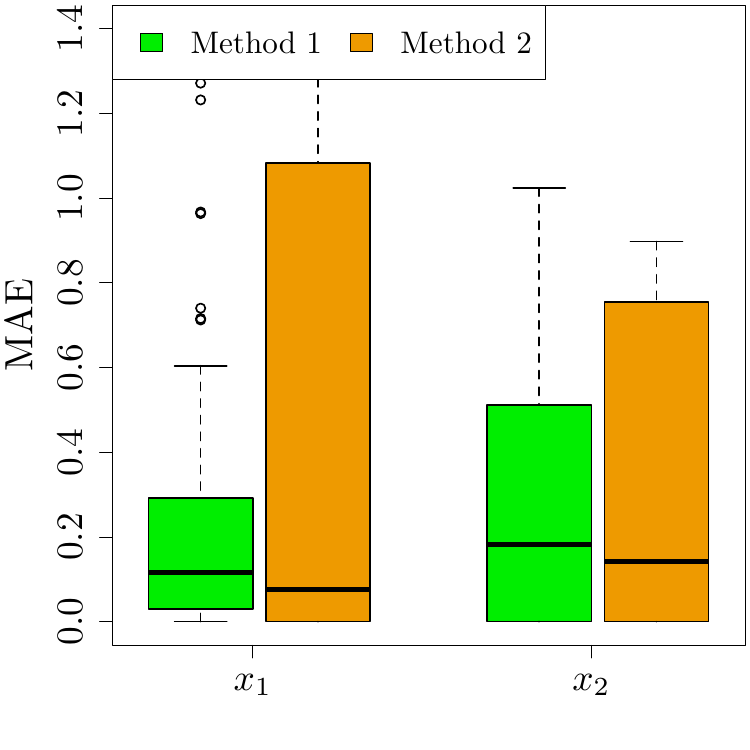}
	\caption{The box plot of MAE and RMS associated with Method 1 (green) and Method 2 (orange) in emulating the van der Pol equation. The two criteria are computed using 100 initial conditions selected randomly from $\lbrack -10, 10\rbrack^2$.  The RMSE and MAE of Method 1 are smaller than those of Method 2.}
	\label{fig:vander_RMSE_MAE}
\end{figure}

\begin{figure}[htpb] 
	\includegraphics[width=0.48\textwidth]{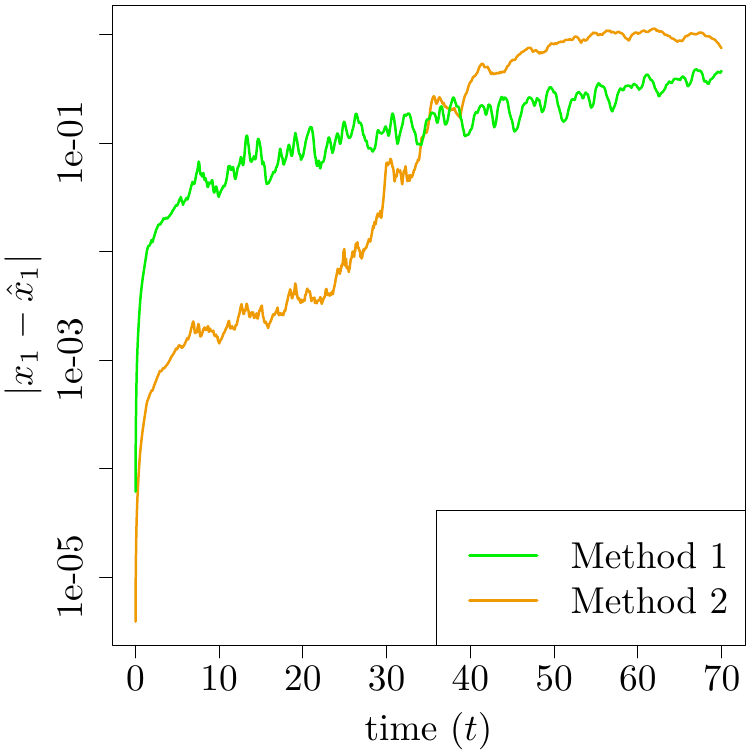}
	\hspace{0.2cm}
	\includegraphics[width=0.48\textwidth]{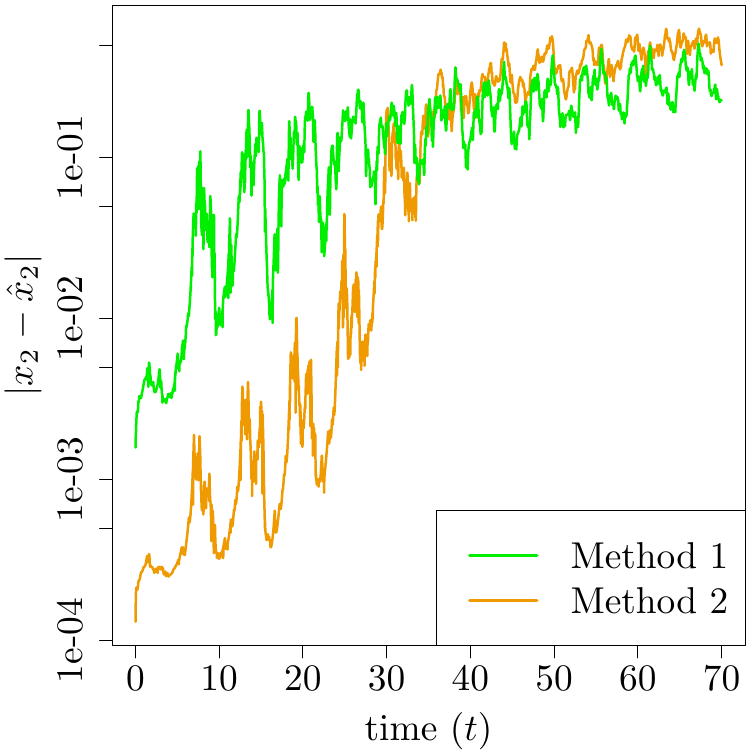}
	\caption{Evolving accuracy of Methods 1 (green) and 2 (orange) over time for $x_1$ (left), and $x_2$ (right). The $y-$axis is on a logarithmic scale.}
	\label{fig:Vander_Accuracy_time}
\end{figure}

\begin{center}
	\begin{tabular}{l || l c}
		& $x_1$ & $x_2$\\ 
		\hline
		Method 1 & 73.3 & 78.4 \\
		\hline
		Method 2 & 31.6 & 34.2
	\end{tabular}
	\newline
\end{center}  
There is a frequency miss-match in the predictions gained by Method 2 after the predictability horizon. Moreover, the amplitude of the predictions (especially $\hat{x}_2$) gradually dampens. However, such issues are less severe in Method 1. This can be confirmed by \Cref{fig:vander_RMSE_MAE} in which the two methods are compared based on MAE and RMS. As can be seen, both criteria obtained by Method 1 are better than those of Method 2 suggesting an improved accuracy in our proposed approach. \Cref{fig:Vander_Accuracy_time} displays the evolving accuracy of both methods over time. Each curve is obtained by computing the mean of $\lvert x_i - \hat{x}_i \rvert$ resulting from 100 initial conditions. While the accuracy trend of Method 2 appears better up to approximately $t = 40$, Method 1 has a more favourable trend thereafter.
%###############################################################################
%=======================================================================
\subsection{Hindmarsh-Rose model}
\label{sec:hindmarsh}
%=======================================================================
The Hindmarsh-Rose (HR) model \cite{hindmarsh1984} is widely used in biology to study the nonlinear dynamics of excitable cells such as neurons. Neurons are specialised cells that are responsible for generating electrical signals called \emph{action potentials}. Information is transmitted via an action potential throughout the nervous system. The HR model is capable of mimicking spiking and bursting which may occur in real cells. The mathematical equations of the HR model are 
\begin{equation}
	\begin{cases}
		\frac{d x_1}{d t} = x_2 - a_1x^3 + a_2x^2 - x_3 + I  \\  \frac{d x_2}{d t} = a_3 - a_4x_1^2 - x_2 \\  \frac{d x_3}{d t} = \varepsilon \left(a_5(x_1 - x_{rest}) - x_3 \right) .
	\end{cases}
	\label{hindmarsh_model}
\end{equation}
The state variable $x_1$ stands for the cell membrane potential and $x_2$ and $x_3$ describe the ionic currents flowing across the membrane through fast and slow ion channels, respectively. The parameter $0 < \varepsilon \ll 1$ is small, which makes $x_3$ a slow variable. $I$ represents the membrane input current and $x_{rest}$ is the rest potential of the system. Having studied a limit cycle and a chaotic behaviour, we then choose in the HR model a complex transient trajectory where the two time scales interplay. The study of transient dynamics is important in many real-world phenomena, see e.g., \cite{goodfellow2012, kittel2017}. To this end, in our experiments the value of the parameters $\varepsilon$, $I$ and $x_{rest} $ are set to 0.01, 2.4 and -1.6, respectively. The typical values for the constant parameters $a_1, \ldots, a_5$ are: $a_1 = 1, a_2 = 2.7, a_3 = 1, a_4 = 5$ and $a_5 = 4$ \cite{barrio2011}. These parameter values are considered in the examples below. 

\begin{figure}[htpb] 
	\includegraphics[width=0.49\textwidth]{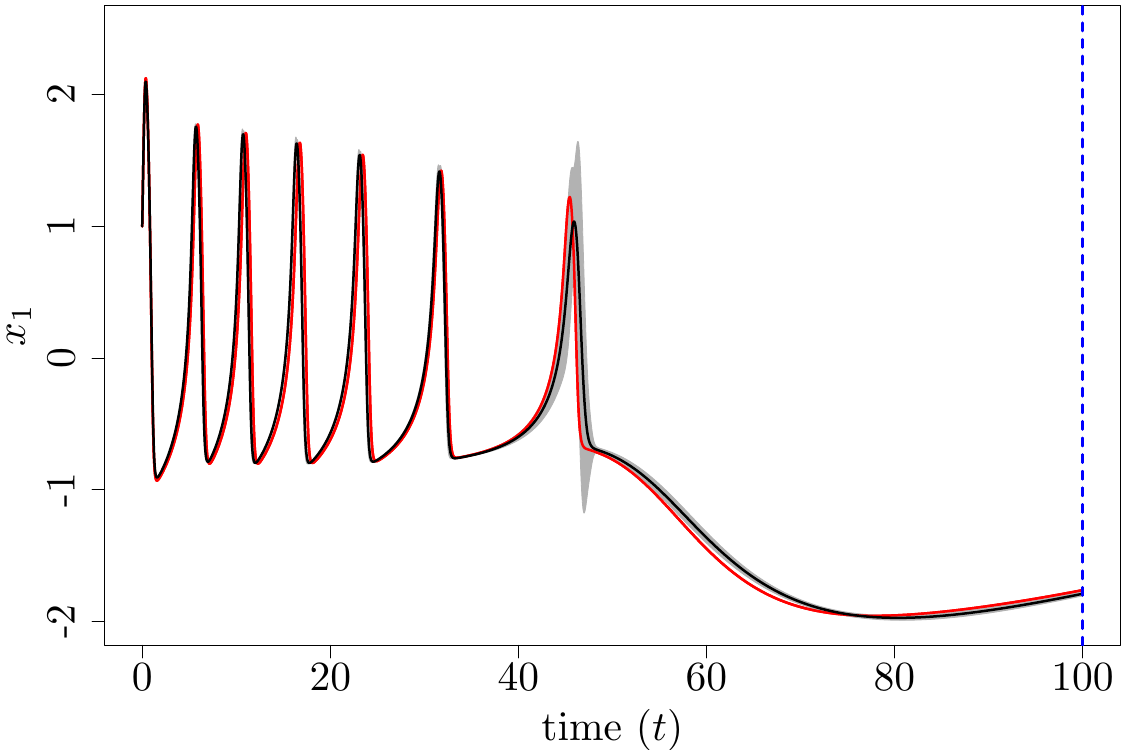}
	\includegraphics[width=0.49\textwidth]{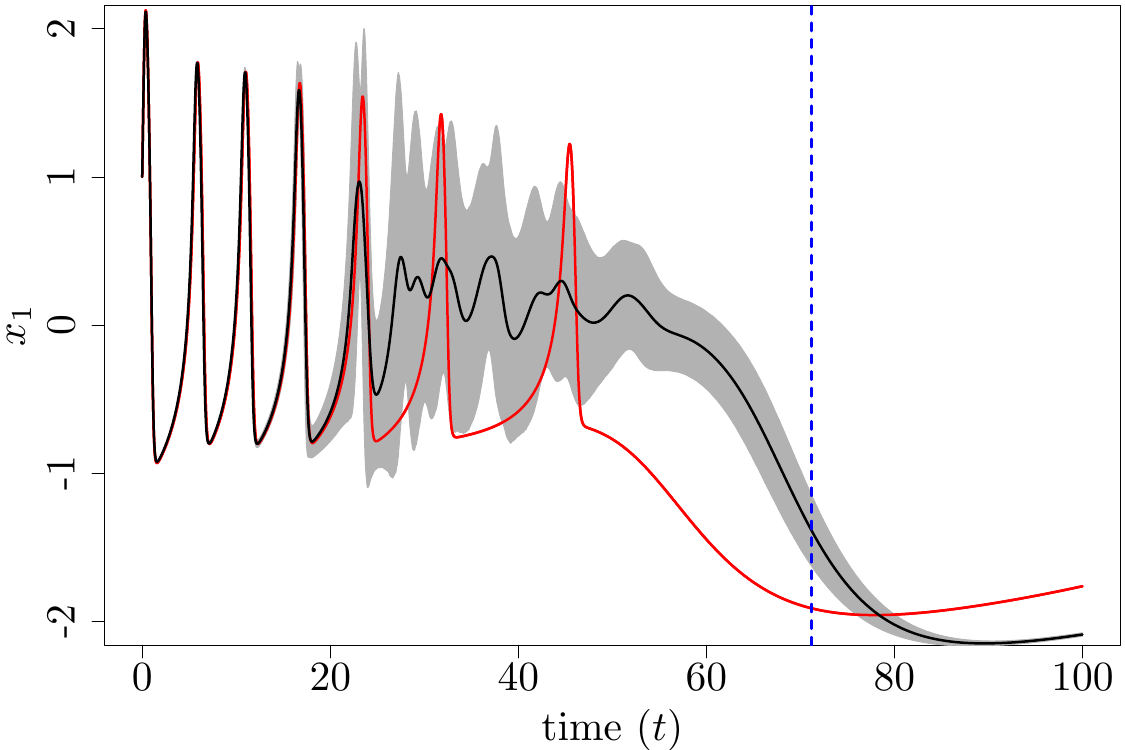}\\
	\includegraphics[width=0.49\textwidth]{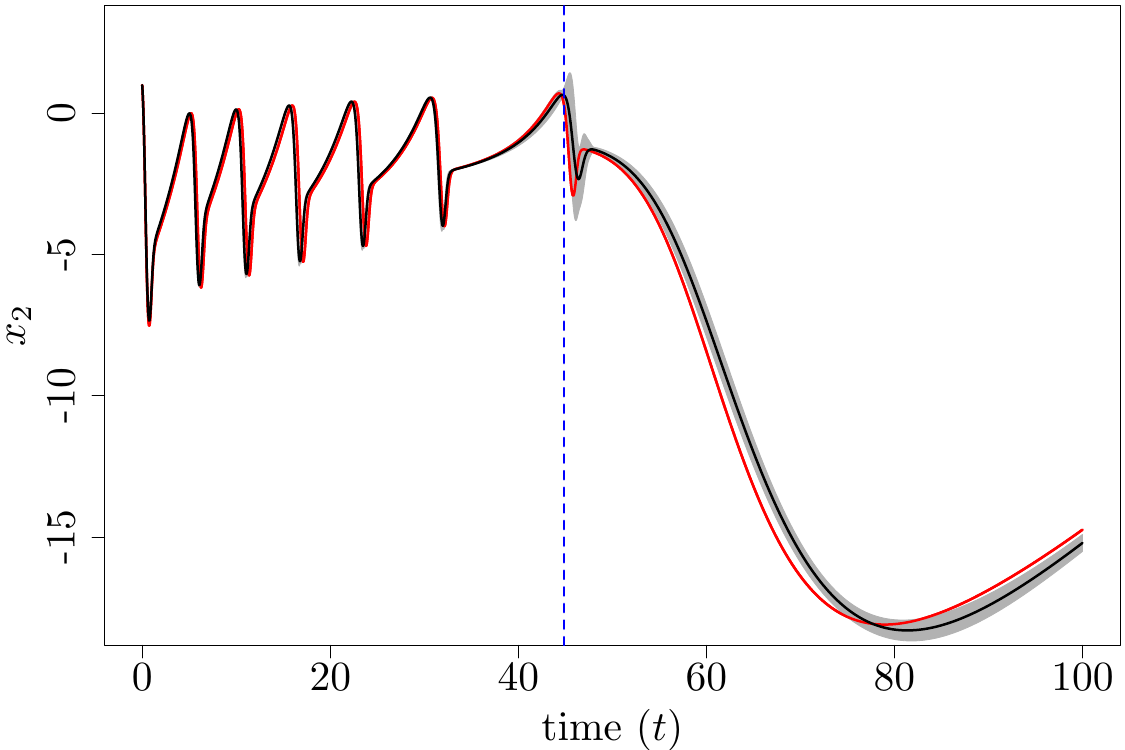}
	\includegraphics[width=0.49\textwidth]{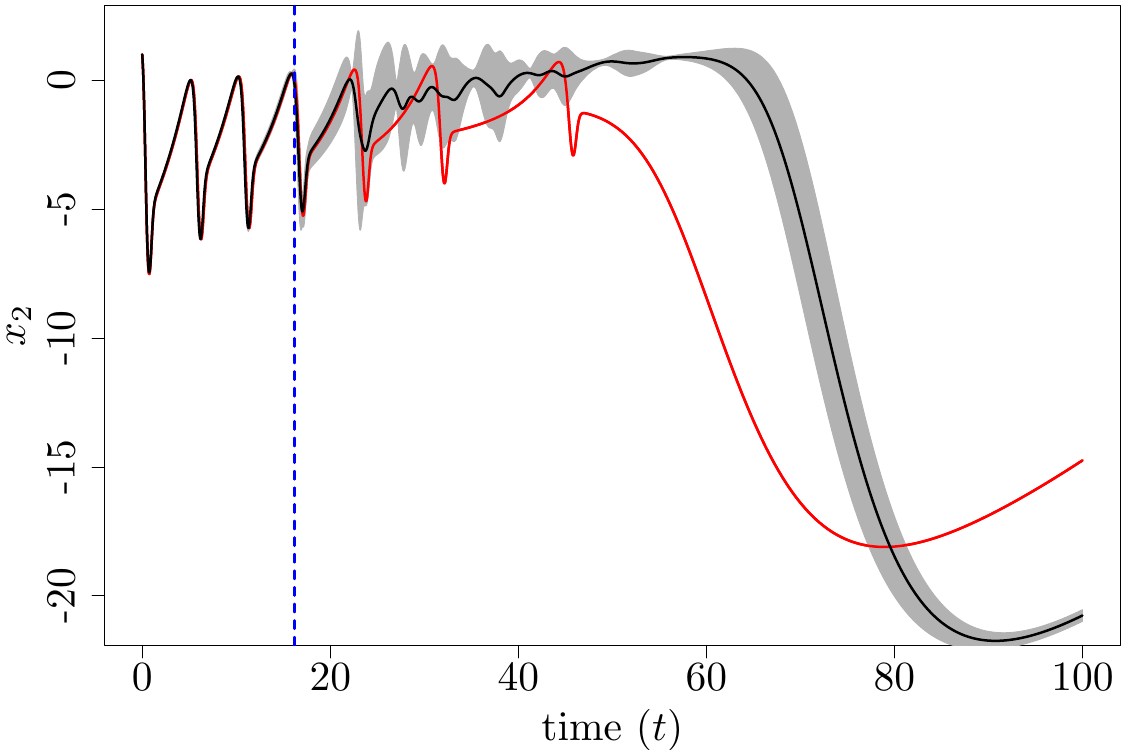}\\
	\includegraphics[width=0.49\textwidth]{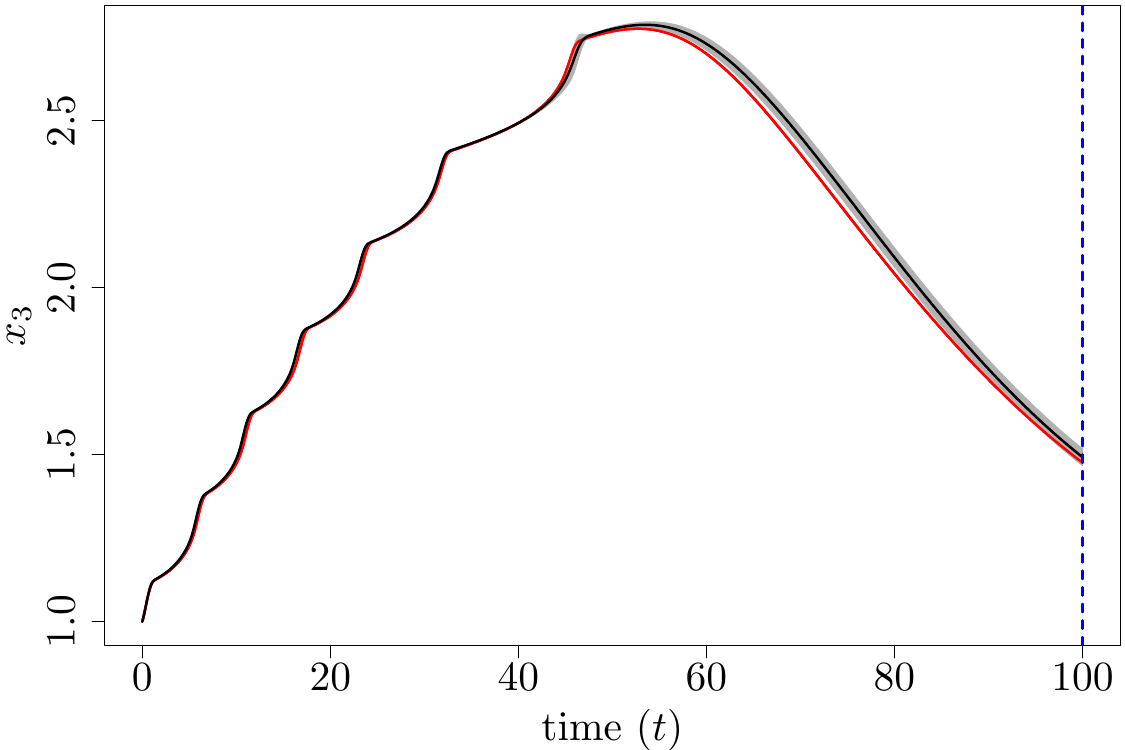}
	\includegraphics[width=0.49\textwidth]{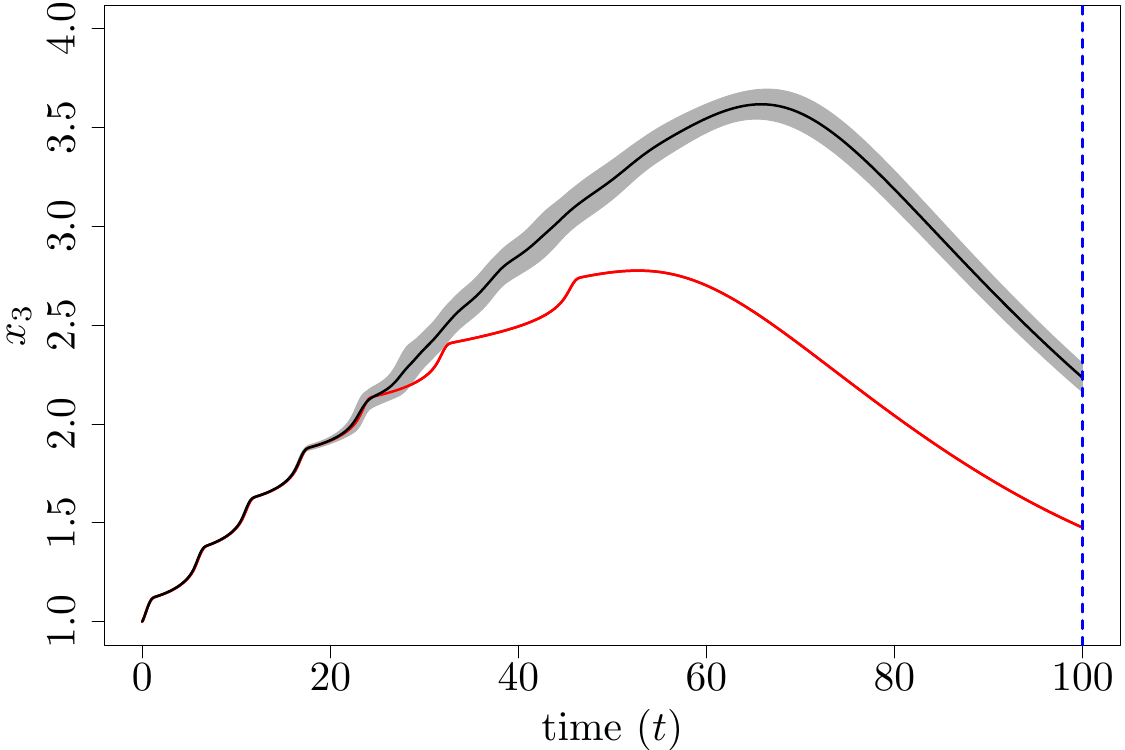}
	\caption{Emulating the HR model with the initial condition $\bx_0 = (1, 1, 1)^\top$ based on Method 1 (left) and Method 2 (right). The proposed approach has a high prediction performance such that the difference between the truth (red) and emulation (black) is negligible. The predictability horizon (dashed blue) occurs at the end of the simulation for $x_1$ and $x_3$.}
	\label{fig:Hindmarsh}
\end{figure}

\begin{figure}[htpb] 
	\includegraphics[width=0.48\textwidth]{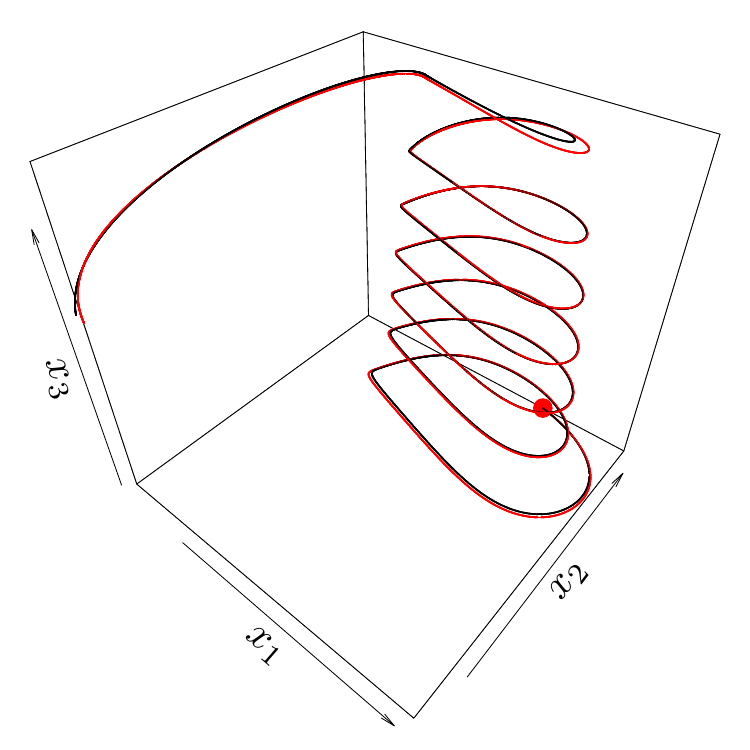}
	\includegraphics[width=0.48\textwidth]{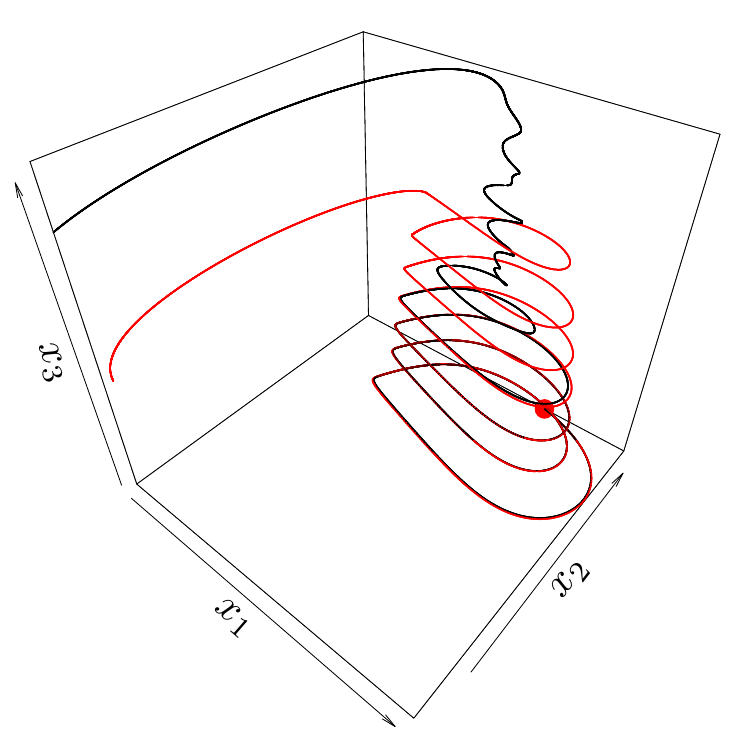}
	\caption{The three-dimensional graph of the HR model based on Method 1 (left) and Method 2 (right) following \Cref{fig:Hindmarsh}. The simulation and emulation are shown in red and black, respectively. The red point indicates the initial condition that is $\bx_0 = (1, 1, 1)^\top$.}
	\label{fig:Hindmarsh_3D}
\end{figure}

The emulation of the HR model obtained by Methods 1 (left) and 2 (right) is presented in Figures \ref{fig:Hindmarsh} and \ref{fig:Hindmarsh_3D}, where the latter shows the corresponding three-dimensional graph. The figures illustrate the HR model (red), its prediction (black), and the associated uncertainty (shaded). The initial condition is $\bx_0 = (1, 1, 1)^\top$ in both cases, and is indicated by the red point in the three-dimensional graph. We see that Method 1 has once again a superior performance compared to Method 2 in emulating all three state variables. In Method 1 the emulator remains reliable until the end of simulation, and the predictability horizon is equal to $100$ in emulating $x_1$ and $x_3$. The coverage probability is computed for the two methods and is summarised below. As can be seen, Method 1 has a better performance in terms of coverage.

\begin{center}
	\begin{tabular}{l || l c r}
		& $x_1$ & $x_2$ & $x_3$\\ 
		\hline
		Method 1 & 62.7 & 58 & 44.4 \\
		\hline
		Method 2 & 42.2 & 42.4 & 30.8
	\end{tabular}
	\newline
\end{center}  

\begin{figure}[htpb] 
	\includegraphics[width=0.48\textwidth]{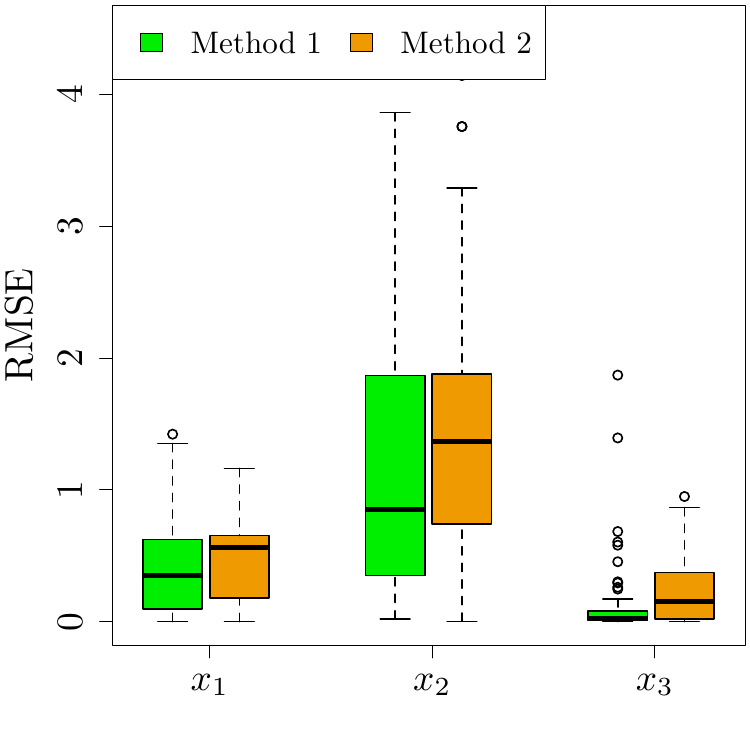}
	\hspace{0.2cm}
	\includegraphics[width=0.48\textwidth]{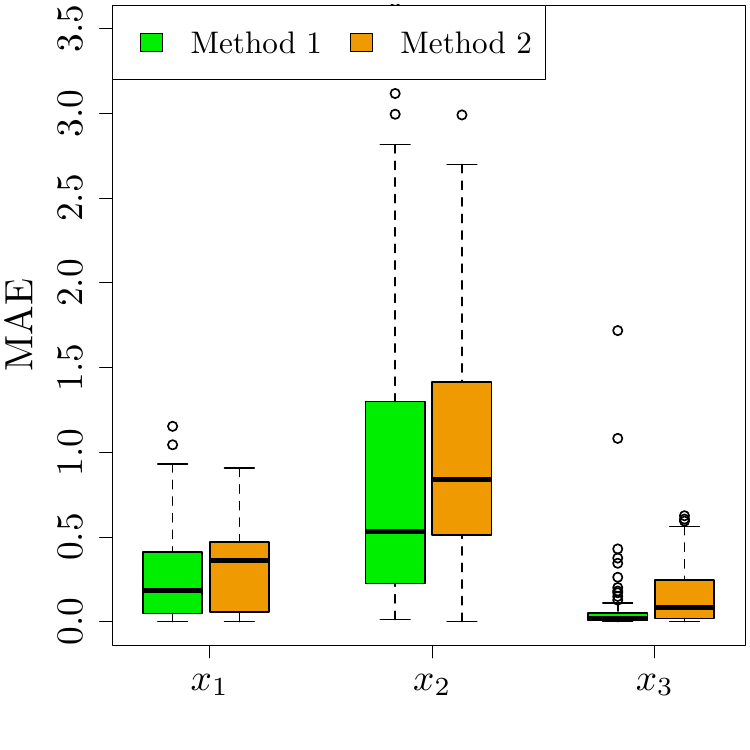}
	\caption{The box plot of MAE and RMS associated with Method 1 (green) and Method 2 (orange) using 100 different initial conditions chosen from $\lbrack -10, 10\rbrack^3$. Method 1 outperforms Method 2 in emulating the HR model.}
	\label{fig:Hindmarsh_RMSE_MAE}
\end{figure}

\begin{figure}[htpb] 
	\includegraphics[width=0.32\textwidth]{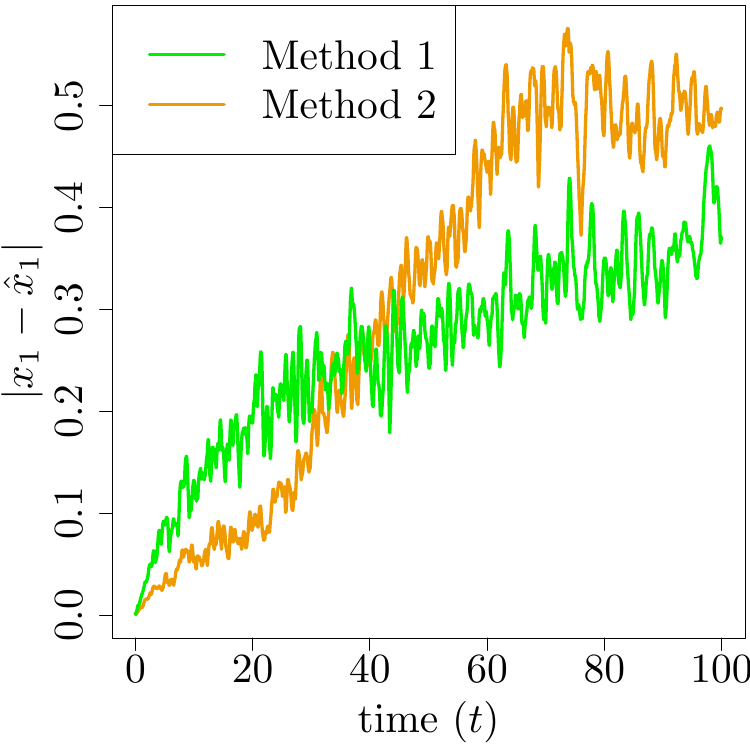}
	\includegraphics[width=0.32\textwidth]{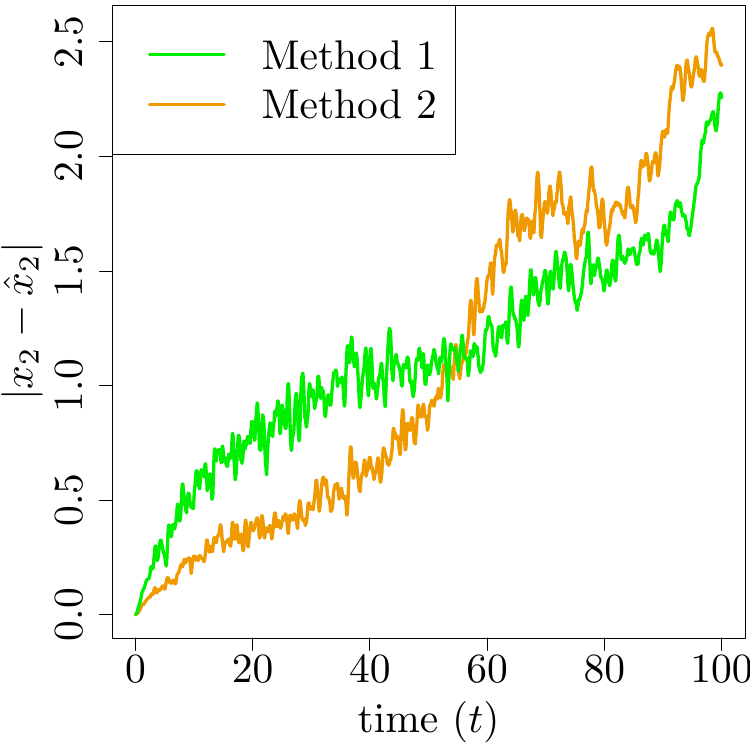}
	\includegraphics[width=0.32\textwidth]{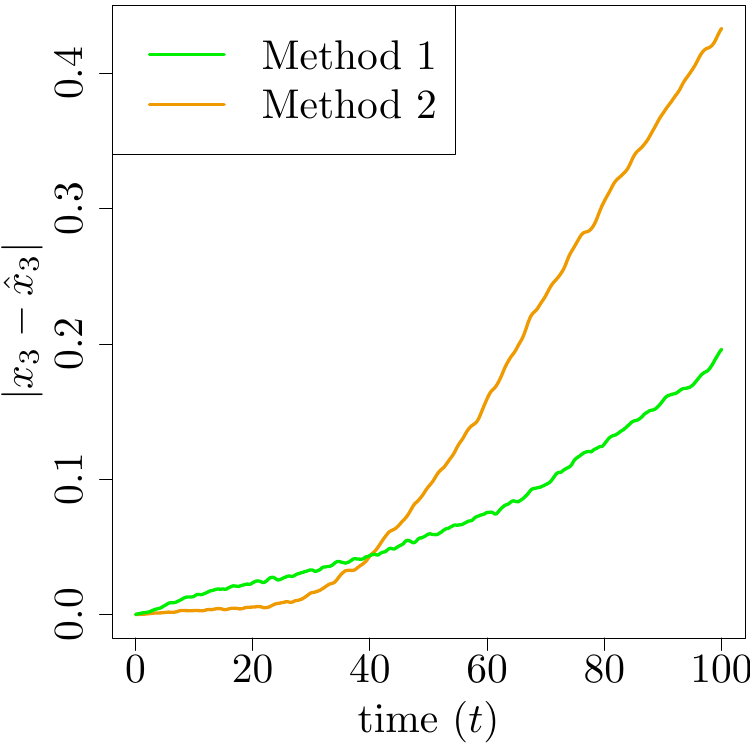}
	\caption{Accuracy trends of Methods 1 (green) and 2 (orange) over time for $x_1$ (left), $x_2$ (middle), and $x_3$ (right). The vertical axis is presented on a logarithmic scale.}
	\label{fig:Hindmarsh_Accuracy_time}
\end{figure}

For further investigation, we compare the two algorithms based on the MAE and RMSE criteria using 100 (random) initial conditions. The box plot of the criteria is demonstrated in \Cref{fig:Hindmarsh_RMSE_MAE}. Furthermore, the accuracy trends of the two methods over time are illustrated in \Cref{fig:Hindmarsh_Accuracy_time}. Each curve represents the average of $\lvert x_i - \hat{x}_i \rvert$ calculated from 100 initial conditions. The results in the previous figures indicate that Method 1 outperforms in emulating the HR model, specially when considering $x_3$.
%###############################################################################
%###############################################################################
%###############################################################################
%===================================================================================================
\section{Discussion}
\label{sec:discussion}
%===================================================================================================
This work presents a novel approach for emulating dynamical simulators, where samples from the posterior GP are defined analytically. In order to do this we approximate the kernel with RFF given that there is no know method to draw exact GP samples. The approximate sample paths are then employed to perform one-step ahead prediction as explained in \Cref{sec:Emul_Dynam_Sys}. We found that the new method performs adequately and can capture a significant portion of the required uncertainty quantification.

In the one-step ahead prediction paradigm uncertainties in emulation will propagate over time. It should be noted that the numerical simulation of a set of ODE (e.g., the numerical simulation of the Lorenz system) also propagates errors which depend upon the numerical scheme employed, as well as properties of the underlying vector field. Even higher-order numerical methods will deviate from the underlying function with time, especially in a chaotic regime such as that we study in the Lorenz system. What we have shown for the systems studied is that the prediction uncertainty increases from step to step up to a predictability horizon that is defined the  time where a change point occurs in the SD of prediction (see \Cref{fig:sd_predict}). After this point the emulator is not able to predict the simulator accurately, however, the uncertainty of the prediction is captured in the proposed method.

\begin{figure}[htpb] 
	\includegraphics[width=0.31\textwidth]{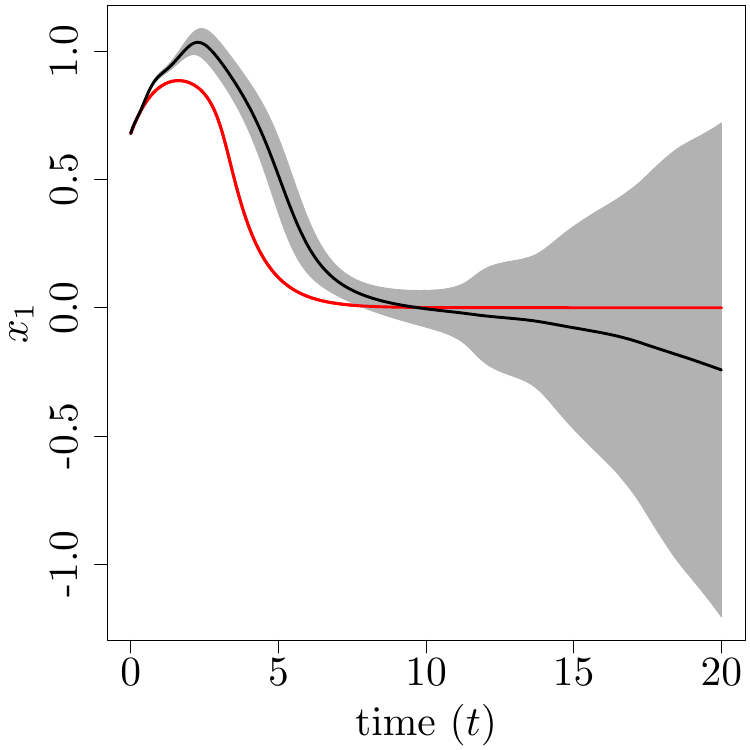}
	\includegraphics[width=0.31\textwidth]{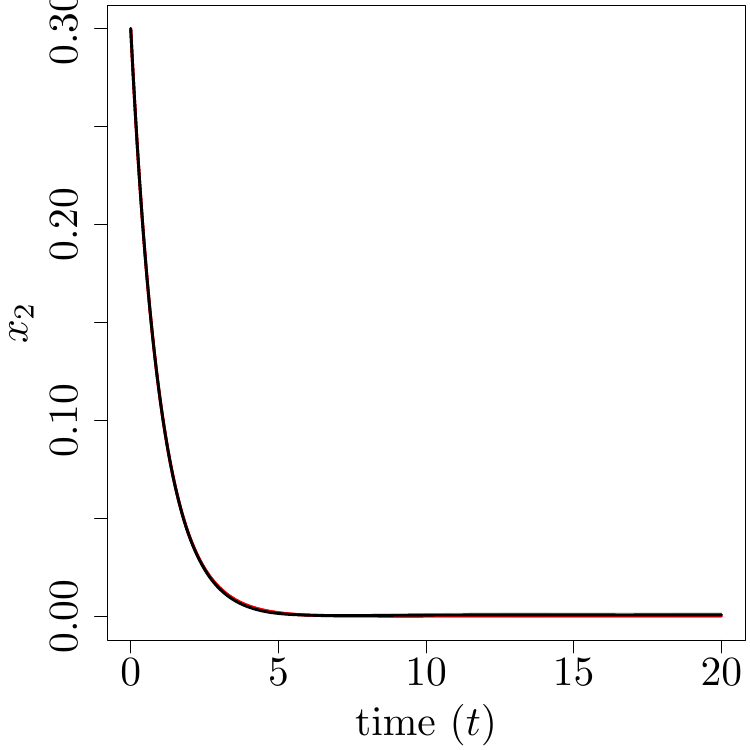}
	\includegraphics[width=0.31\textwidth]{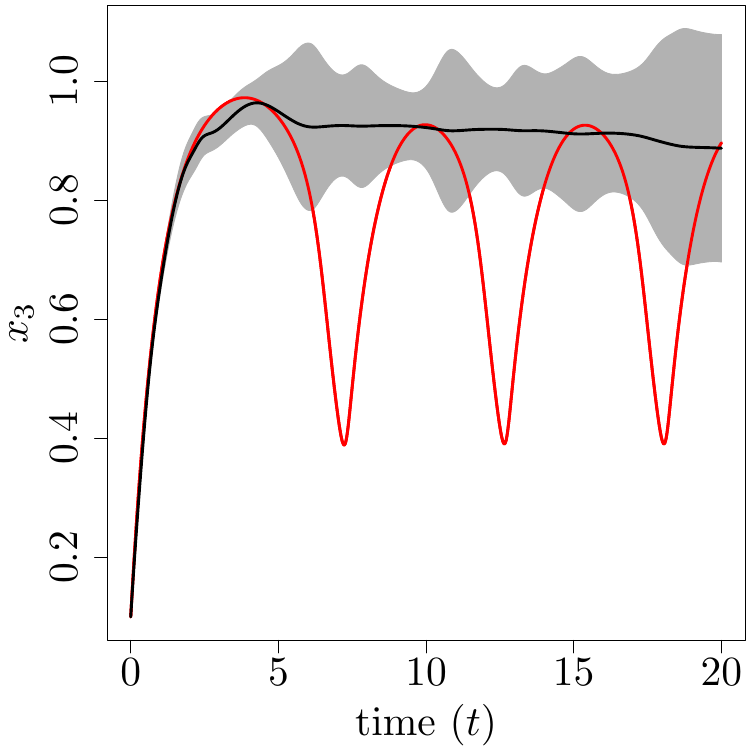}\\
	\includegraphics[width=0.31\textwidth]{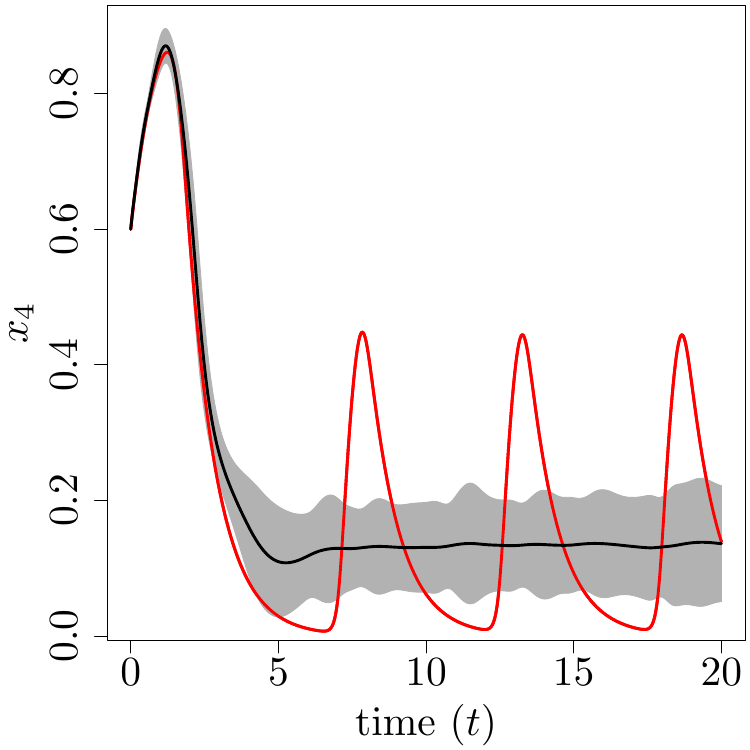}
	\includegraphics[width=0.31\textwidth]{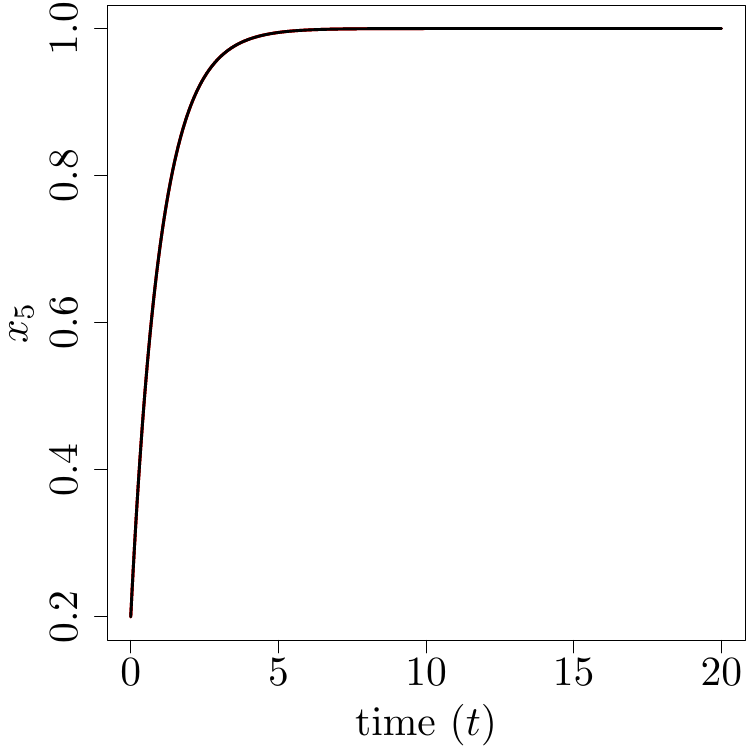}
	\includegraphics[width=0.31\textwidth]{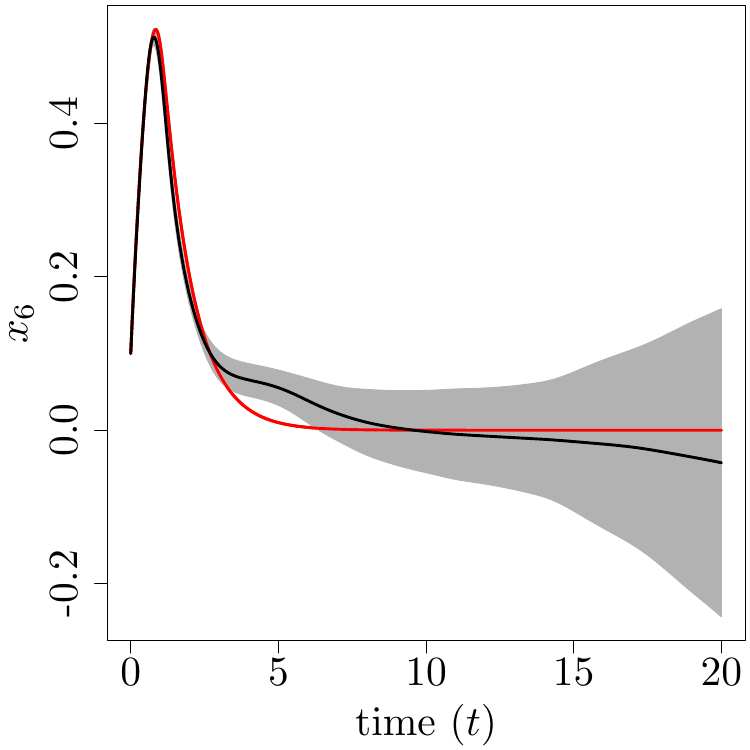}
	\caption{The prediction (black) and associated uncertainty (shaded) in emulating the six-dimensional dynamic model (red) given in \Cref{eq:6D_example}. The prediction is highly accurate for the second and the fifth state variable, and most of the variation of $x_1$ and $x_6$ is captured.}
	\label{fig:6D_example}
\end{figure}

The dimensionality of the dynamical systems we considered in this work is two or three. The applicability of our proposed methodology to higher dimensional problems needs more investigations though. As an example, we have applied our method to a six-dimensional model whose equations are given below
\begin{align}
	\begin{cases}
		\frac{d x_1}{d t} &= \frac{1}{1 + e^{-10(W_{11}x_1 + W_{12}x_2 + \ldots + W_{16}x_6 - a_1)}} - x_1\\ \; \vdots &= \hspace{2.2cm} \vdots \\  \frac{d x_6}{d t}  &= \frac{1}{1 + e^{-10(W_{61}x_1 + W_{62}x_2 + \ldots + W_{66}x_6 - a_6)}} - x_6  ,
	\end{cases}
	\label{eq:6D_example}
\end{align}
with $a_1 = a_2 = a_3 = 0.8$ and  $a_4 = a_5 = a_6 = -0.8$. $W_{ij}$ is the $ij$-th element of the following regulatory matrix
\begin{equation*}
	W = 
	\begin{pmatrix}
		1 & 2 & 0 & 0 & 0 & 0\\
		-2 & 1 & 0 & 0 & 0 & 0\\
		0 & 0 & 1 & 2 & 0 & 0\\
		0 & 0 & -2 & 1 & 0 & 0\\
		0 & 0 & 0 & 0 & 1 & 2\\
		0 & 0 & 0 & 0 & -2 & 1
	\end{pmatrix} .
\end{equation*}
Such an ODE system is used in various domains including gene regulatory networks \cite{ogorelova2020}. \Cref{fig:6D_example} shows the results. As can be seen, we have a high prediction accuracy for the state variables $x_2$ and $x_5$. The emulator can capture most of the variation of the first and the sixth state variable.  In the other cases, the prediction tends to the average of the process when the emulator is not able to predict the simulator well.
%###############################################################################
%###############################################################################
%===================================================================================================
\section{Conclusions}
\label{sec:conclusion}
%===================================================================================================
We proposed a novel data-driven approach for emulating deterministic complex dynamical systems implemented as computer codes. The output of such models is a time series and presents the evolving state of a physical phenomenon over time. Our method is based on emulating the short-time numerical flow map of the system and using draws of the emulated flow map in an iterative manner to perform one-step ahead predictions. The flow map is a function that returns the solution of a dynamic system at a certain time point, given initial conditions. In this paper, the numerical flow map is emulated via a GP and its approximate sample paths are generated with random Fourier features. The approximate GP draws are employed in the one-step ahead prediction paradigm which results in a distribution over the time series. The mean and variance of that distribution serve as the time series prediction and the associated uncertainty, respectively. The proposed method is tested on several nonlinear dynamic simulators such as the Lorenz, van der Pol, and Hindmarsh-Rose models. The results suggest that our approach can emulate those systems accurately and the prediction uncertainty can capture the true trajectory with a good accuracy. A future work direction is to conduct quantitative studies such as uncertainty quantification and sensitivity analysis on computationally expensive dynamical simulators emulated by the method suggested in this paper. 
%===================================================================================================
\section*{Acknowledgments}
HM and PC would like to thank the Alan Turing Institute for funding this work. 
%=======================================================================
\begin{appendices}
\section{Reproducing kernel Hilbert space}
\label{appendix:rkhs}
%===================================================================================================
Let $\mathcal{H}$ be a Hilbert space of functions defined on $\mathcal{X}$. The function $k(\cdot, \cdot)$ is called a reproducing kernel of $\mathcal{H}$, and $\mathcal{H}$ is an RKHS, if it satisfies
\begin{enumerate}
	\item $\forall  \bx \in \mathcal{X} \rightarrow k(\bx, \cdot) \in \mathcal{H} $ ; and
	\item $\forall \bx \in \mathcal{X}$ and $\forall f \in \mathcal{H} , \,  \langle f, k(\cdot, \bx) \rangle_ \mathcal{H}= f(\bx)$  (the reproducing property) .
\end{enumerate}
For the reproducing kernel $k$ and the feature map $\phi : \bx \mapsto k(\bx, \cdot)$, we have
\begin{equation}
	\left\langle \phi(\bx), \phi(\bx^\prime) \right\rangle_{\mathcal{H}} =  \left\langle k(\bx, \cdot), k(\bx^\prime, \cdot) \right\rangle_{\mathcal{H}}= k(\bx, \bx^\prime) \, , ~~ \forall \bx, \bx^\prime \in \mathcal{X} ,
	\label{reprod_prop}
\end{equation} 
which follows directly from the reproducing property. The above equation suggests that the input space $\mathcal{X}$ can be projected into a higher (or infinite) dimensional feature space (through $\phi$) where the learning procedure can be more successful. 
\end{appendices}
%=======================================================================
\bibliography{biblio}
\bibliographystyle{plain}
\end{document}